\documentclass[%
 reprint,
superscriptaddress,
 amsmath,amssymb,
 aps,pre,hyphens,floatfix
]{revtex4-2}

\usepackage[]{amsmath}
\usepackage{amssymb}
\usepackage{enumerate}
\usepackage[usenames,dvipsnames,svgnames,table]{xcolor}
\usepackage{IEEEtrantools}
\usepackage{graphicx}% Include figure files
\usepackage{array}
\usepackage{multirow}
\usepackage{wrapfig}
\usepackage{fullpage}
\usepackage{overpic}
\usepackage{verbatim}
\usepackage{float}
\usepackage{neuralnetwork}
\usepackage[caption=false]{subfig}
\usepackage{multirow}
\usepackage{diagbox}

\usepackage[normalem]{ulem}  % without normalem the long lines in the bibliography do not break up!

\newcommand{\tcr}[1]{\textcolor{black}{ #1}}

\begin{document}

\preprint{APS/123-QED}

\title{Transfer learning of phase transitions in percolation and directed percolation}
% Force line breaks with \\
% \thanks{A footnote to the article title}%
%\author{Complexity Science}
%%\email[]{sjm@mails.ccnu.edu.com}
%\affiliation{Key Laboratory of Quark and Lepton Physics (MOE) and Institute of Particle Physics, Central China Normal University, Wuhan 430079, China}

\author{Jianmin Shen}
%\email[]{sjm@mails.ccnu.edu.com}
\affiliation{Key Laboratory of Quark and Lepton Physics (MOE) and Institute of Particle Physics, Central China Normal University, Wuhan 430079, China}

\author{Feiyi Liu}
\email[]{fyliu@mails.ccnu.edu.com} 
\affiliation{Key Laboratory of Quark and Lepton Physics (MOE) and Institute of Particle Physics, Central China Normal University, Wuhan 430079, China}
\affiliation{Institute for Physics,E{\"o}tv{\"o}s Lor\'and University\\1/A P\'azm\'any P. S\'et\'any, H-1117, Budapest, Hungary
}

\author{Shiyang Chen}
\email[]{shiyang-chen@mails.ccnu.edu.com}
\affiliation{Key Laboratory of Quark and Lepton Physics (MOE) and Institute of Particle Physics, Central China Normal University, Wuhan 430079, China}

\author{Dian Xu}
%\email[]{dianxu.work@mails.ccnu.edu.com}
\affiliation{Key Laboratory of Quark and Lepton Physics (MOE) and Institute of Particle Physics, Central China Normal University, Wuhan 430079, China}

\author{Xiangna Chen}
%\email[]{xiangna@mails.ccnu.edu.com}
\affiliation{Key Laboratory of Quark and Lepton Physics (MOE) and Institute of Particle Physics, Central China Normal University, Wuhan 430079, China}

\author{Shengfeng Deng}
%\email[]{steven@gmail.com}
\affiliation{Institute of Technical Physics and Materials Science, Center for Energy Research, Budapest 1121, Hungary}

\author{Wei Li}
%\email[]{liw@mail.ccnu.edu.cn}
\affiliation{Key Laboratory of Quark and Lepton Physics (MOE) and Institute of Particle Physics, Central China Normal University, Wuhan 430079, China}
\affiliation{Max-Planck-Institute for Mathematics in the Sciences, 04103 Leipzig, Germany}

\author{G\'abor Papp}
%\email[]{gabor.papp@ttk.elte.hu}
\affiliation{Institute for Physics,E{\"o}tv{\"o}s Lor\'and University\\1/A P\'azm\'any P. S\'et\'any, H-1117, Budapest, Hungary
}

\author{Chunbin Yang}
%\email[]{cbyang.ccnu.edu.com}
\affiliation{Key Laboratory of Quark and Lepton Physics (MOE) and Institute of Particle Physics, Central China Normal University, Wuhan 430079, China}

\date{\today}
% It is always \today, today, but any date may be explicitly specified

\begin{abstract}

The latest advances of statistical physics have shown remarkable performance of machine learning in identifying phase transitions. In this paper, we apply domain adversarial neural network (DANN) based on transfer learning to studying non-equilibrium and equilibrium phase transition models, which are percolation model and directed percolation (DP) model, respectively. With the DANN, only a small fraction of input configurations (2d images) needs to be labeled, which is automatically chosen, in order to capture the critical point. To learn the DP model, the method is refined by an iterative procedure in determining the critical point, which is a prerequisite for the data collapse in calculating the critical exponent $\nu_{\perp}$. We then apply the DANN to a two-dimensional site percolation with configurations filtered to include only the largest cluster which may contain the information related to the order parameter. The DANN learning of both models yields reliable results which are comparable to the ones from Monte Carlo simulations. Our study also shows that the DANN can achieve quite high accuracy at much lower cost, compared to the supervised learning.  

% BrainPainter\footnote{Source code: \url{https://github.com/mrazvan22/brain-coloring}} is customisable, easy to use, and can run straight from the web browser: \url{http://brainpainter.csail.mit.edu}.
\end{abstract}
% diagram showing the aim: input numbers and output images
\maketitle

\section{Introduction}
\label{intro}

In the age of artificial intelligence, machine learning (ML) \cite{jordan2015machine,goodfellow2016machine} has promptly become a significant means of scientific research. Due to its great power in image recognition and feature extracting, a huge amount of complex data can be analysed handily in statistical physics \cite{engel2001statistical,mehta2014exact,mehta2019high,carleo2019machine,carrasquilla2020machine},  which provides us a new choice to deal with phase transitions beyond conventional field theory methods \cite{domb1996critical,amit2005field} and Monte Carlo simulations \cite{hammersley2013monte}.

Recently, supervised learning \cite{carrasquilla2017machine,van2017learning,zhang2019machine} and unsupervised learning \cite{wang2016discovering,shen2021supervised,wetzel2017unsupervised,hu2017discovering,wang2017machine,wang2021unsupervised} methods have been applied to learning problems that involve phase transitions. The difference between the two is that the former requires labeled data as input while the latter does not, but the goals of both are to train models with better compatibility. In the field of phase transition, supervised learning mainly identifies or classifies phases of matter. While unsupervised learning techniques, such as principal component analysis (PCA) \cite{pearson1901liii,abdi2010principal}, stochastic neighbor embedding (T-SNE) \cite{van2008visualizing,van2014accelerating} and autoencoder \cite{bourlard1988auto,hinton1994autoencoders,hinton2006reducing,shen2021machine} etc, are more suitable for clustering and dimensionality reduction. At present, a well performed technique named transfer learning (TL) \cite{kouw2018introduction,xu2020transfer,
ch2017machine,huembeli2018identifying,malo2019applications} mixing both labeled and unlabeled data, has been popular in dealing with images. It can not only obtain the critical exponent of the phase transition model through data collapse like supervised learning does, but extract the feature representation from the original data like unsupervised learning does. Additionally, TL has been used in community or module detection \cite{eaton2012spin,zhang2014phase}, which may also lead to phase transitions. In reality data are rarely labeled since feature engineering with vast amount of data is time-consuming and computationally expensive. This fact thus gives rise to the emergence of TL, which has lately become a crucial branch of ML. The basic idea of TL is to enable the model to translate unlabeled data in target domain into labeled data in source domain. In Ref. \cite{ajakan2014domain} it has been shown that the domain adversarial neural network (DANN)  has better performance than traditional neural networks (NN) and support vector machines (SVM) do. In Ref. \cite{huembeli2018identifying}, the DANN has also successfully identified paradigmatic phase transition models, including the Ising model, the Bose-Hubbard model, and the Su-Schrieffer-Heeger model with disorder, which opens the door for the study of many-body localization problem. We are therefore intrigued by the extent of effectiveness of DANN in traditional phase transition models, especially of non-equilibrium, which is the main motivation of this work.     
The two adopted models are directed percolation (DP) model and percolation model. These two models have broad applicability, as well as uniqueness of properties that have to be paid attention. The DP model \cite{hinrichsen2000non,lubeck2004universal} represents the most important universality class of non-equilibrium phase transitions that we can understand so far \cite{hinrichsen2000non,lubeck2004universal,odor2004,henkel2008non,haken1975cooperative}, the DP universality class. Quite a few non-equilibrium models were confirmed to belong to the DP class. DP was seldom studied by machine learning method \cite{shen2021supervised}, not to mention the DANN method.  Its machine learning may enable us to understand a wide range of non-equilibrium systems which may fall into the same universality class as DP does. Such studies may also help to unveil the mystery of university class of non-equilibrium phase transitions, whose understanding is still at the very beginning. The percolation model \cite{essam1980percolation,christensen2005complexity} is a special one of equilibrium phase transition \cite{amit2005field,domb1996critical}, whose order parameter is not the particle density as in most models. Its machine learning may shed light on more elaborate models which have similar characteristics of order parameters as the percolation does. \tcr{Reference \cite{huembeli2018identifying} has} successfully detected the underlying phase transitions and predicted the transition points in the Bose-Hubbard model (system of interacting Bose gas lattice model which displays a transition between super-fluids and normal fluids) and the Su-Schrieffer-Heeger model (system of interacting spinless fermions which displays a transition from topological trivia phase to a nontrivial phase). Here in this work, we can not only extract essential information embedded at the critical regime using the DANN method, namely predict the transition point, but obtain fairly accurate correlation exponents. Our study expands the regime on which DANN is capable of working. To improve the accuracy of measurements of critical points and critical exponents, a method to extend source domain is developed to expand our labeled set. Additionally, we examine when we expect the DANN to work properly.

The main structure of this paper is as follows. In Sec.~\ref{Models}, the two phase transitions models of interest will be given. Sec.~\ref{method} gives the method of adversarial domain adaptation. Sec.~\ref{DANN_result} includes the data sets and DANN learning results of the phase transitions models employed. In Sec.~\ref{Discussions}, we discuss our major findings.  Sec.~\ref{Conclusion} is a summary of this work.

\section{Transfer learning of phase transitions}
\subsection{Models}\label{Models}

In the following we briefly introduce two models, a non-equilibrium directed bond percolation in 1+1 dimensions, and an equilibrium two-dimensional site percolation.

\subsubsection{The DP model}

Compared to equilibrium phase transitions, non-equilibrium ones possess an extra dimension, namely time. In this part, we focus on a class of phase transitions that involve absorbing states, where the time evolution stops ~\cite{hinrichsen2000non,lubeck2004universal,jensen1993nonequilibrium,rossi2000universality} and the system is stuck in inactive states. The physical nature of absorbing phase transitions is a competitive mechanism of proliferation and annihilation of quantities that can be particles, energy, molecules, viruses, and so on and so forth. Here, we first consider the DP model with only one absorbing state. The DP model represents one of the most significant class of non-equilibrium phase transitions, the DP universality class. Lattice models like contact process (CP), Domany-Kinzel (DK) model, and pair-contact process (PCP), all belong to the DP universality class.

\begin{figure*}[!htb]
%\begin{tabular}{cc}
\begin{tabular}{ccc}   
%    \rotatebox{90}{$\xleftarrow[]{time \quad  t}$}
    \includegraphics[width=0.4\textwidth]{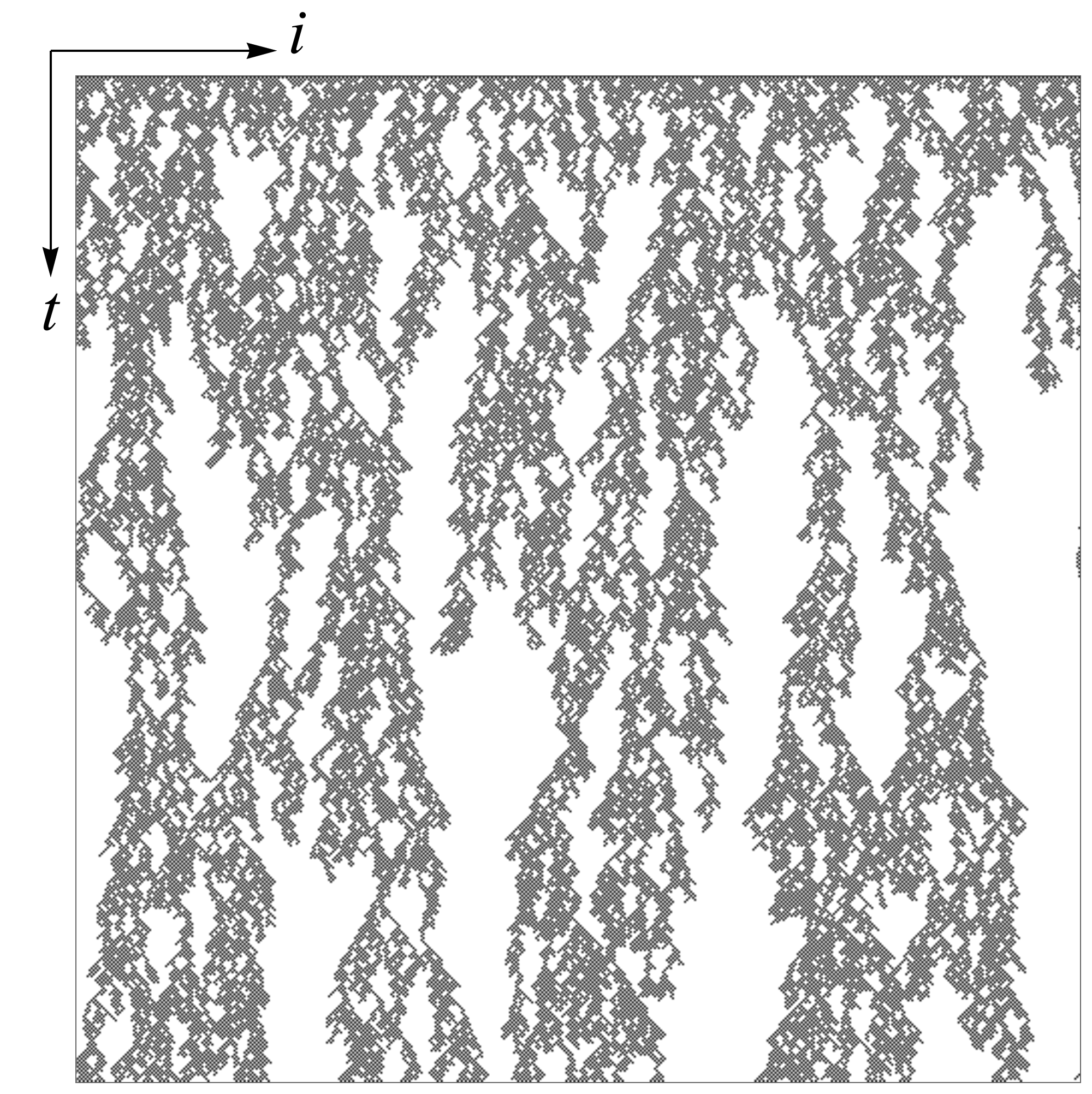} &
    \includegraphics[width=0.4\textwidth]{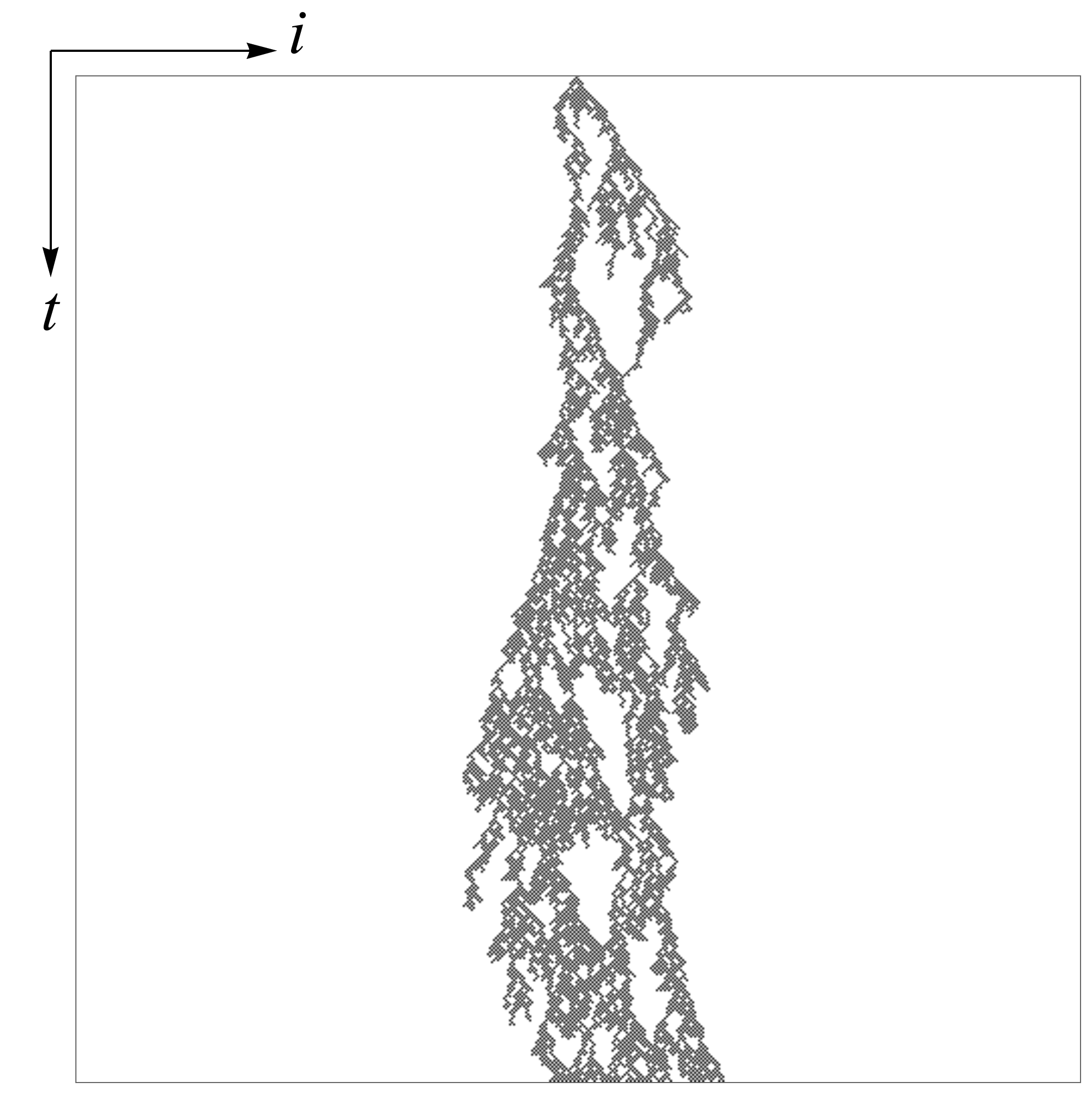} \\
     Fully occupied lattice &  Single active seed
\end{tabular} 
%\put(25,1){\color{red}{hkkjk}}      
\caption{The critical configurations of bond DP in (1+1) dimensions, starting from a fully occupied lattice (left panel) and from a single active seed (right panel), respectively, where $L = 500$, the time step is $500$ and the bond probability $p$ is $0.6447$.}
\label{DP_configuration}
\end{figure*}

In the simplest version of bond DP, the evolution is done on a tilted (1+1) dimensional square lattice, and a bond is formed at the time step with probability $p$ from an existing bond (see Fig.~\ref{DP_configuration}). Here, we are using periodic boundary conditions. This model may be interpreted as a reaction-diffusion process of interacting particles:
representing the active particle as $A$ and the empty site as $\varnothing$, the reaction-diffusion mechanism of DP is
\begin{equation}
\label{DP-diff}
  \left\{
\begin{array}{rcl}
 \mbox{self-destruction:} &   A \longrightarrow \varnothing,  \\
 \mbox{diffusion:} &   \varnothing + A \longrightarrow A + \varnothing, \\
 \mbox{offspring production:} &  A \longrightarrow 2A, \\
 \mbox{coagulation:} & 2A \longrightarrow A. \\ 
\end{array} \right.
\end{equation}

ML techniques can process complex data produced by various lattice models of non-equilibrium phase transitions. These models include unitary or binary random reaction processes, which may contain diffusion or non-diffusion motions. Since within the universality class only the critical exponents are common and the position of the phase transition may vary, we fix the model to the simplest case mentioned above. Fig.~\ref{DP_configuration} shows DP's configurations generated from two different initial conditions, including a fully occupied lattice (left panel) and a single active seed (right panel). In this paper we use the former one (the reason was explained in one of our previous articles \cite{shen2021supervised}) to generate configurations with different bond percolation probabilities. In the case of fully occupied lattice, all sites are occupied at the origin of (1+1)-dimensional bond DP, while only one site is occupied in the case of a single active seed. The DP model evolves according to following rules,
\begin{equation}
  s_{i}(t+1)=\left\{
\begin{array}{rcl}
1     &      & {if \quad s_{i-1}(t) = 1 \quad and \quad z_{i}^{- } < p,} \\
1     &      & {if \quad s_{i+1}(t) = 1 \quad and \quad z_{i}^{+ } < p,} \\
0     &      & {otherwise,}
\end{array} \right.
\end{equation}
where $i$ means a spatial coordinate, $t$ a discrete time variable, and $p$ the bond probability of a connection between two adjacent sites. A binary variable $s_{i}(t)$ represents the state of site $i$ at time  $t$, $z_{i}^{\pm} \in (0, 1) $ is a random number taken from a uniform distribution. Therefore states of all sites determine the configuration of the system.

Likewise, the universality class of non-equilibrium ones is also determined by their critical exponents. DP has three independent critical exponents $\beta$, $\nu_{\perp}$, and $\nu_{\parallel}$. For the (1+1)-dimensional bond DP, close to the critical point, the order parameter can be expressed by the steady-state density~\cite{hinrichsen2000non},
\begin{equation}
 \rho_{a}(p)\widetilde{\propto}(p-p_{c})^{\beta},
\end{equation}
where $\rho_{a}$, $p$, and $p_{c}$ denote particle density, bond probability, and critical bond probability, respectively. The steady-state of (1+1)-dimensional bond DP also obeys following power-laws, 
\begin{equation}
  \xi_{\perp}\sim \mid p-p_{c}\vert^{-\nu_{\perp}},   \quad  \xi_{\parallel}\sim \mid p-p_{c}\vert^{-\nu_{\parallel}},
\end{equation}
where $\xi_{\perp}$ denotes the spatial correlation length and $\nu_{\perp}$ is the spatial correlation exponent, $\xi_{\parallel}$ is the temporal correlation length and $\nu_{\parallel}$ is the temporal correlation exponent. Since the density is encoded in the full configuration, we provide configurations exemplified in the left panel of Fig.~\ref{DP_configuration} as an input to the DANN. Some care should be taken with the choice of the characteristic temporal length $t_c$, in order to achieve stable results. As shown in~\cite{hinrichsen2000non}, due to scaling the characteristic time $t_c$ is proportional to $L^{z/d}$, where $z=1.580(1)$ is the dynamical exponent and $d$, the spatial dimension of the percolation which is 1 here.
%except the the time step $T = 3L$.

\subsubsection{The Percolation model}

\begin{figure*}[!thb]
\centering

%\vspace{1.6ex}

%\def\myFigureScale{0.2}
\def\myFigureScale{0.2}
%\hspace{0.8ex}
\includegraphics[scale=\myFigureScale]{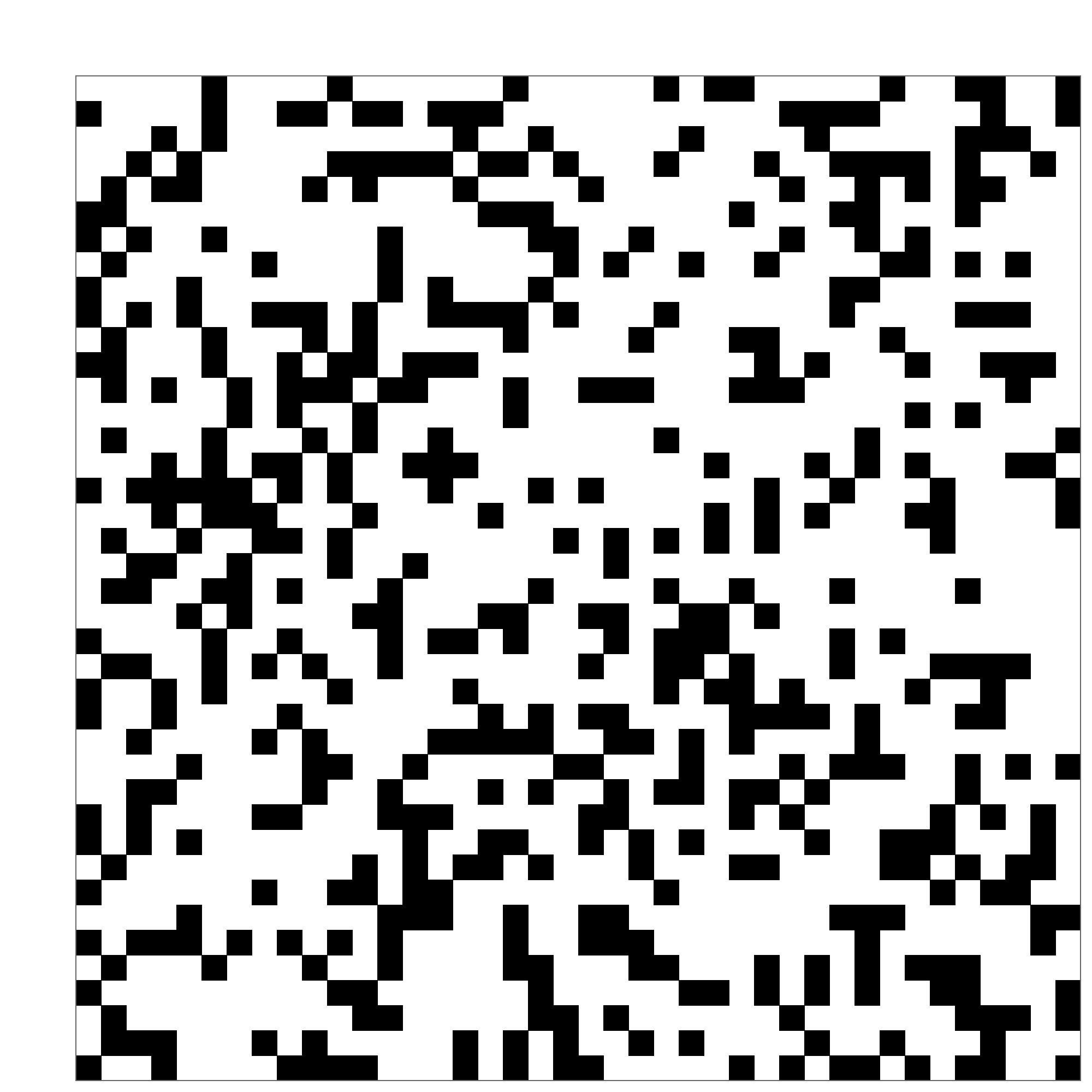}
\includegraphics[scale=\myFigureScale]{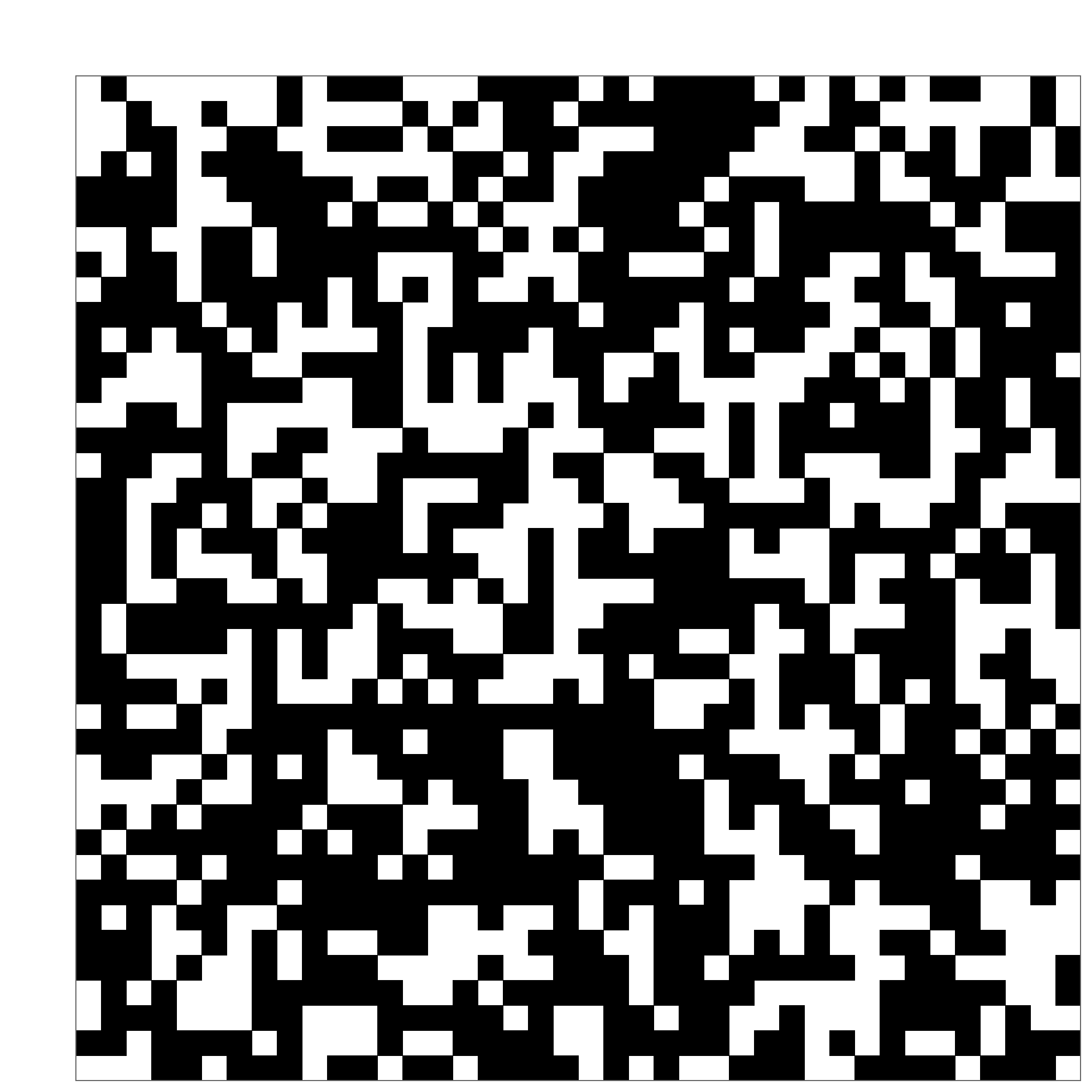}
\includegraphics[scale=\myFigureScale]{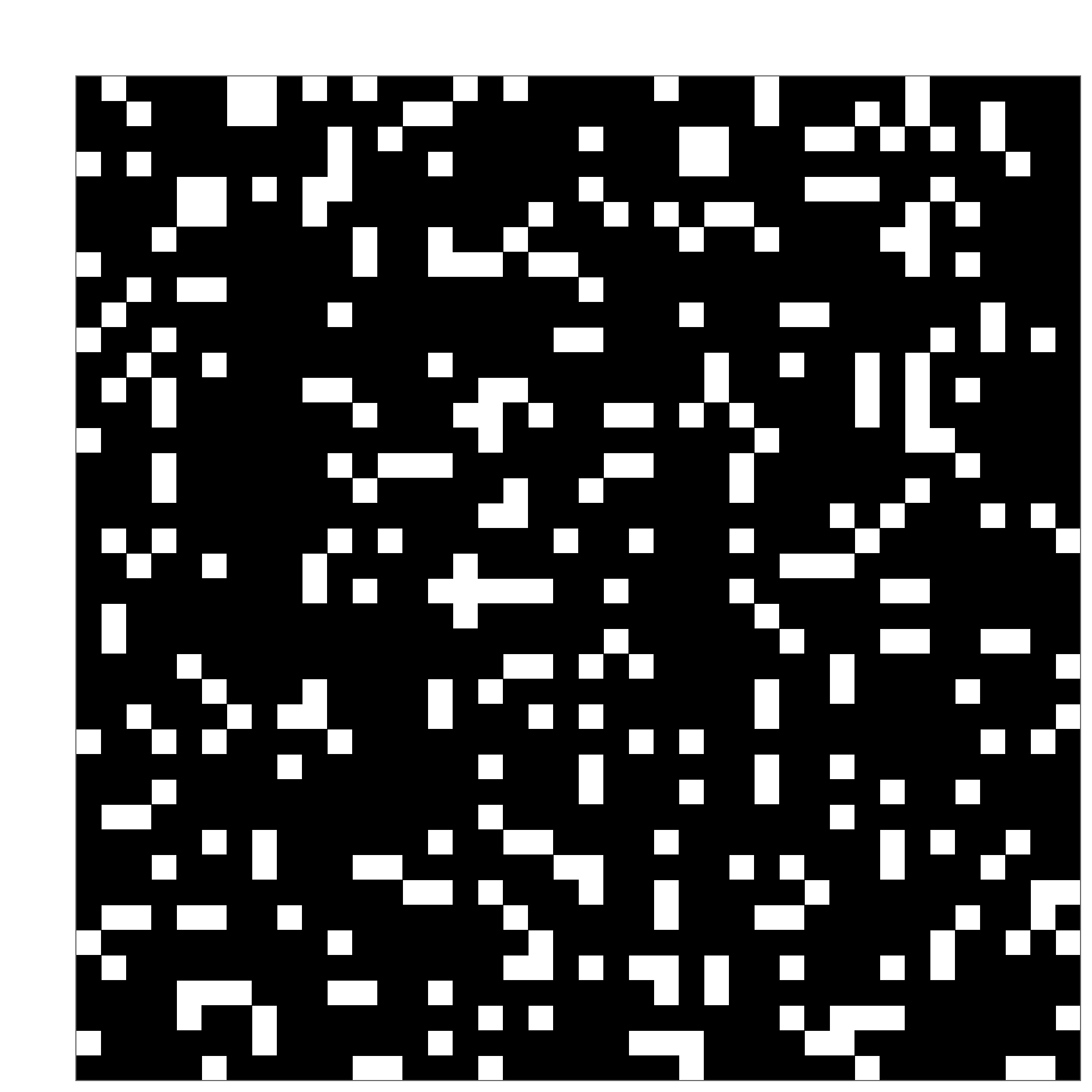}

%\hspace{0.8ex}
\includegraphics[scale=\myFigureScale]{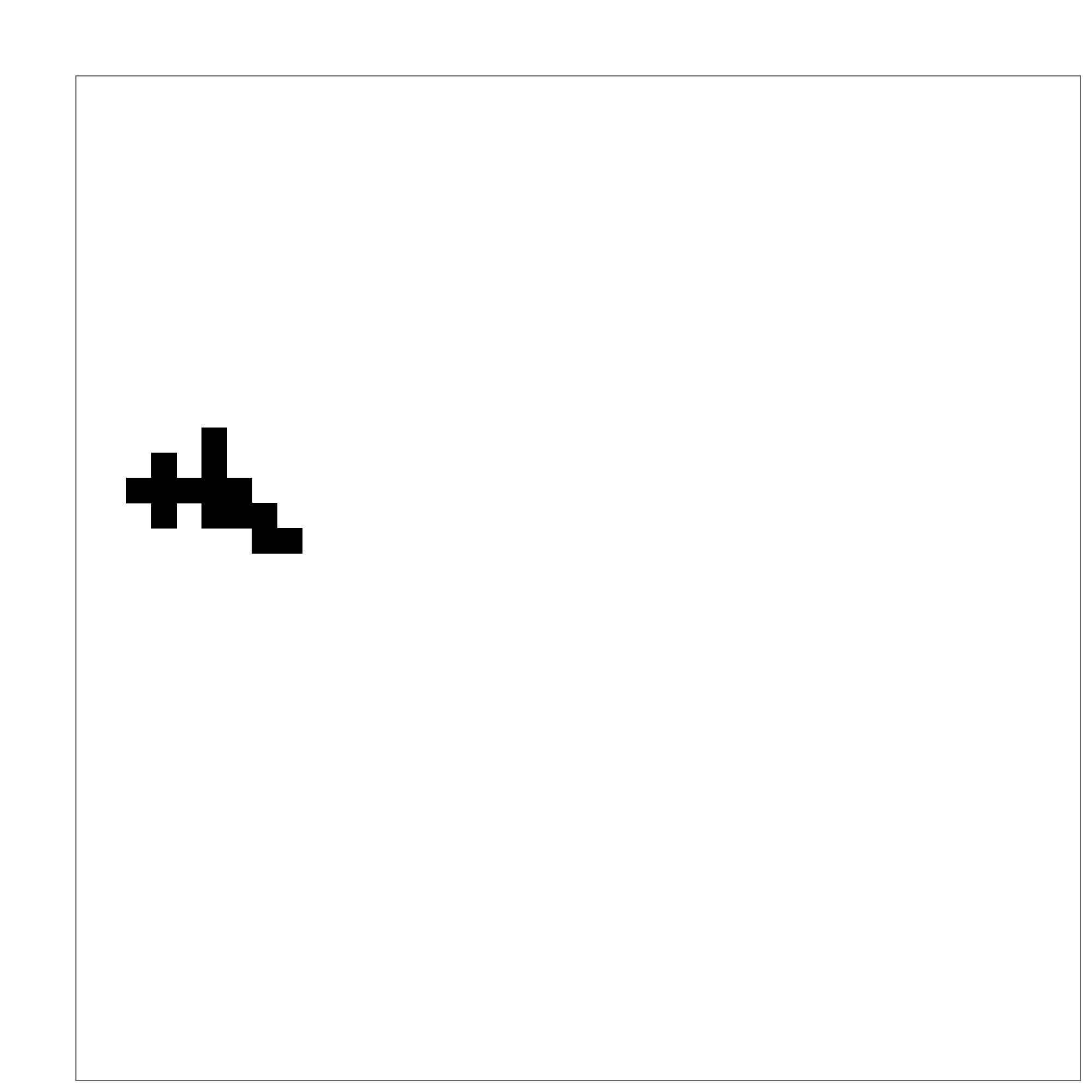}
\includegraphics[scale=\myFigureScale]{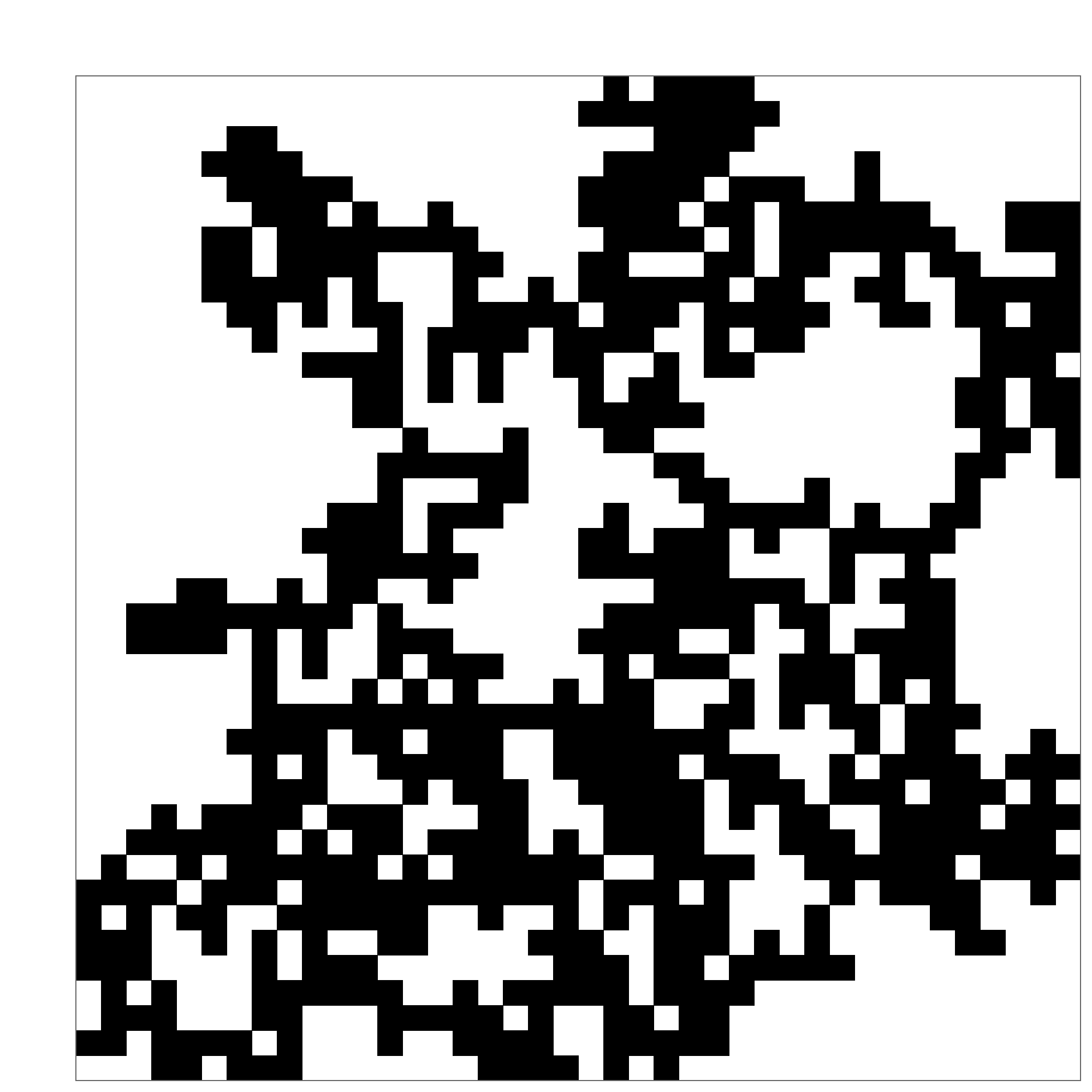}
\includegraphics[scale=\myFigureScale]{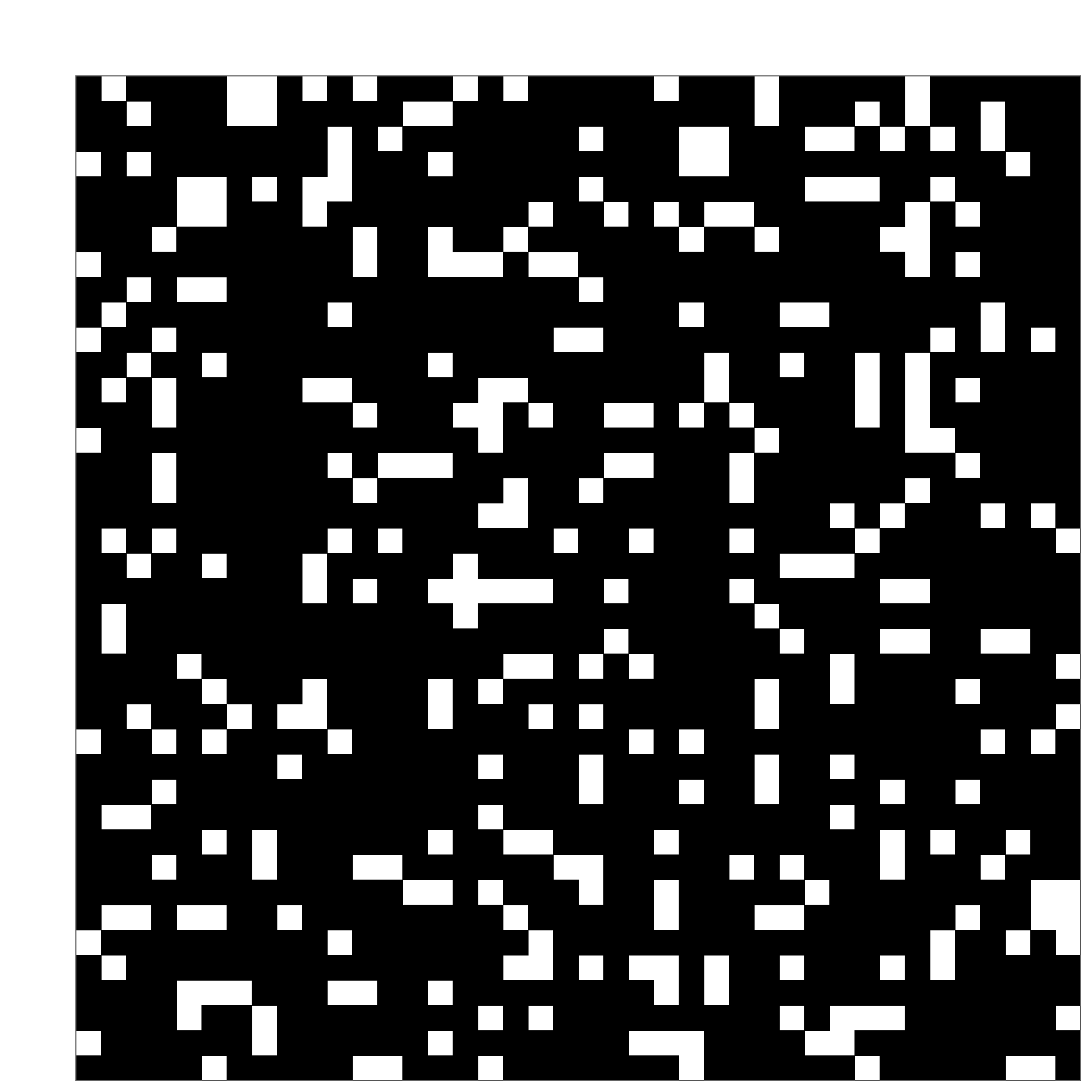}
\caption{The top panels are raw configurations of two-dimensional site percolation generated by probabilities of $0.3$, $0.593$, and $0.8$ respectively, where lattice size is $40$. The bottom panels are corresponding, largest clusters generated from the top panels.}
\label{percolation_configuration}
\end{figure*}

The second model for our study is the two-dimensional site percolation with periodic boundary conditions, where the probability of a site being occupied is controlled by the parameter $p$. We can simply characterise its occupation mechanism by the following equations,
\begin{equation}
 s_{ij}=\left\{
\begin{array}{rcl}
1     &      & {if \quad z_{ij} < p,} \\
0     &      & {otherwise,}
\end{array} \right. 
\end{equation}
where $ij$ means the spatial coordinates of a two-dimensional system, $p$ is the occupation probability, and $z_{ij}$ is a random number taken from a uniform distribution between 0 and 1. Then the occupancy status of all lattice sites determines the configuration corresponding to a given occupation probability. A few typical configurations are shown in the upper panels of Fig.~\ref{percolation_configuration} at different occupation probabilities.

The order parameter of the site percolation is the fraction of the sites $P_\infty(p)$  in the lattice belonging to the infinite cluster (for infinite lattice) or, for finite lattices belonging to the incipient infinite clusters or simply percolating clusters. An infinite system is percolating if only one cluster is infinite, while in finite systems the percolating cluster emerges as the occupied sites form a channel from top to down or left to right. The fraction $P_\infty(p)$ can be also identified with a probability of a site belonging to the percolating cluster, and around the critical point it scales with the critical exponent $\beta$~\cite{hinrichsen2000non} as,
\begin{equation}
P_{\infty}(p) \propto (p - p_{c})^{\beta} \quad \mbox{for} \quad p \rightarrow p_{c}^{+} \,.
\end{equation}

Now, the order parameter is only related to the largest cluster. Hence, using DANN for predicting the transition point, contrary to the DP case where the full configuration is provided as the input, here for percolation learning only the largest cluster is singled out as input. Namely, for each configuration only the sites belonging to the largest cluster are left, as shown in the lower panels of Fig.~\ref{percolation_configuration}.
For two-dimensional (square lattice) site percolation, the critical occupation value $p_{c} = 0.592746$ \cite{christensen2005complexity}. 

\subsection{The domain adversarial neural network (DANN) method}\label{method}
\begin{figure*}[!thb]
\centering
 
\includegraphics[width=1\textwidth]{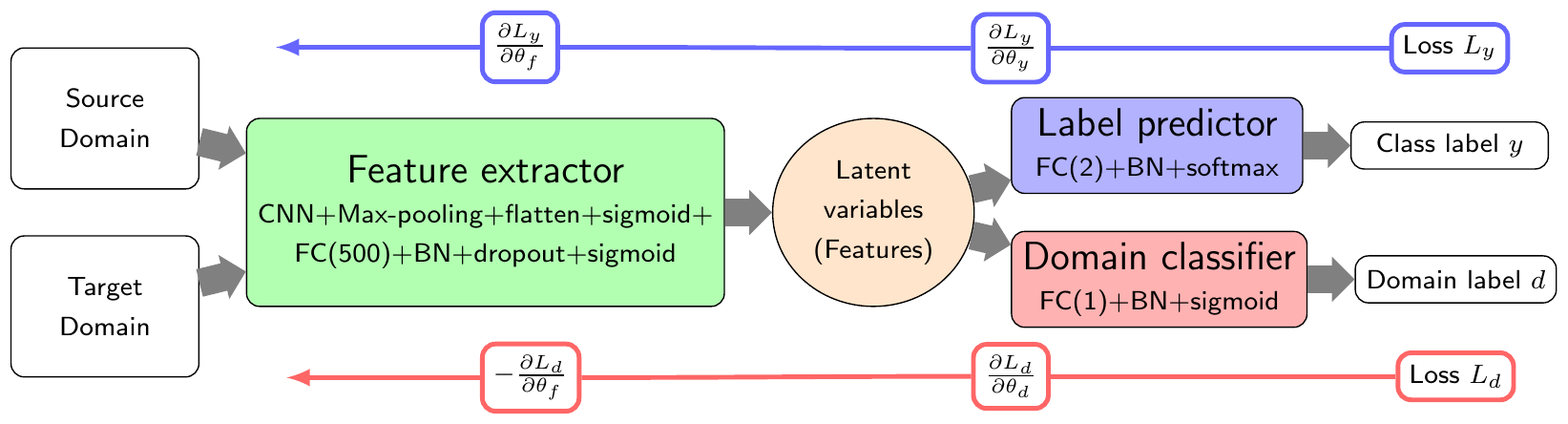}    
\caption{The neural network schematic structure of adversarial domain adaptation. $\theta_f, \theta_d, \theta_y$ are the parameters for the feature extractor, domain classifier and label predictor, respectively. In the training of the feature extractor (\tcr{green in color figure}) the normal feedback propagation is inverted for the classifier parameters in order to create classification independent features. The `Feature extractor' is shown in Fig.~\ref{feature_extractor}.}
\label{domain}
\end{figure*}
\begin{figure*}[!thb]
\centering 
\includegraphics[width=0.9\textwidth]{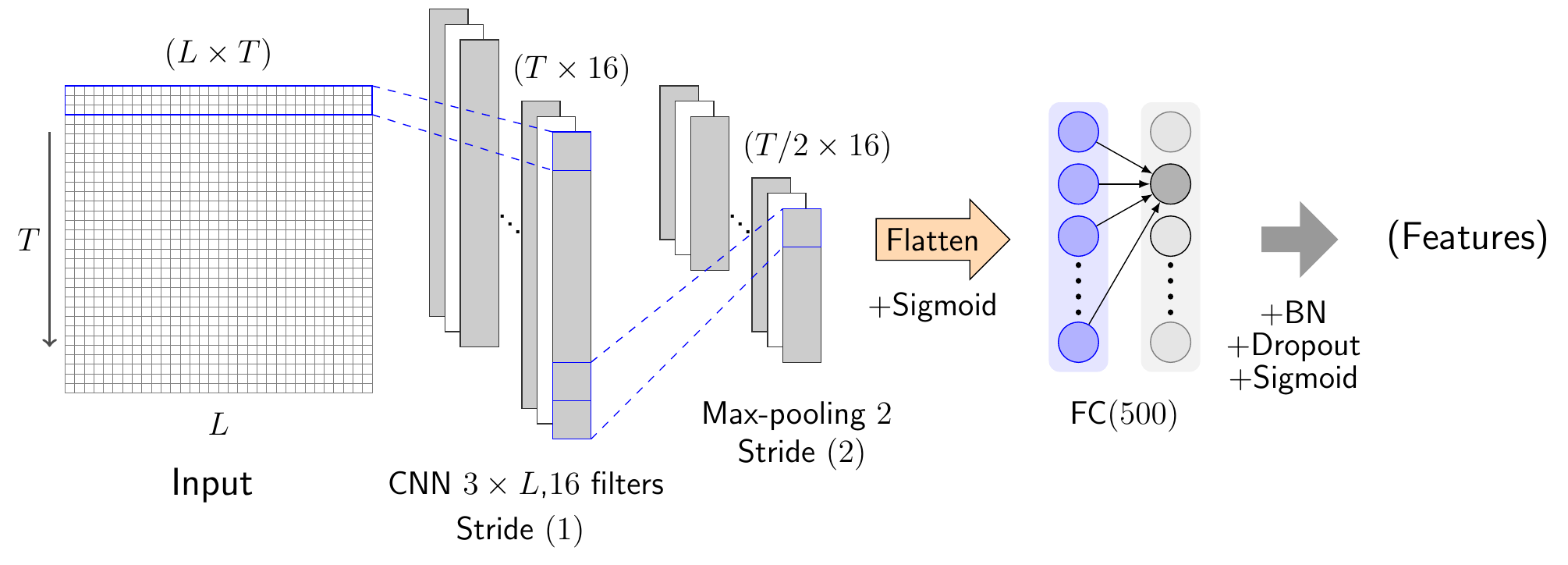}    
\caption{The structure of feature extractor for the input of a two-dimensional system of size $L \times T$. }
\label{feature_extractor}
\end{figure*}

To find out the parameters of the phase transitions, we construct a DANN, and feed it with the configurations as in Fig.~\ref{DP_configuration} for (1+1)- dimensional directed bond percolation and as in the lower part of Fig.~\ref{percolation_configuration} for two-dimensional site percolation. The architecture of the network is the same for the two models which are trained, however, separately.

The major obstacle for applying the neural networks, most of which are density based training methods, to phase transitions for physics models is the lacking of labeled data near critical points, which may cause the supervised network prediction of critical points to deviate from the genuine results~\cite{ganin2016domain}. We choose here DANN, because it is expected to predict the critical points more accurately compared to the traditional perception network~\cite{ajakan2014domain}. A performance comparison between the traditional NN and DANN will be given in Sec.~\ref{DANN and supervised learning}. The detailed mechanism of DANN approach is written as follows.

The basic components of DANN are feature extractor, label predictor and domain classifier, as shown in Fig.~\ref{domain}. The prepared training dataset is divided into two parts: a source domain and a target domain. The source domain consists of labeled data $x_s$ and its label $y_s$ with distribution $\mathcal{S} = \{(x_s ,y_s )\}$, and the target domain contains all the unlabeled data $x_t$ only with distribution $\mathcal{T} = \{(x_t)\}$. The goal of DANN is to predict the labels $y_t$ of the unlabeled configurations $x_t$ through finding its similarities with the labeled dataset $\{x_s\}$. For the purpose, domain label $d_i$ is introduced to distinguish data $x_i\in\{x_s\}\cup\{x_t\}$ from source domain ($d_i=0$ if $x_i\in \mathcal{S}$) or target domain  ($d_i=1$ if $x_i\in \mathcal{T}$). Then, the input $x_i\in\mathcal{S}\cup\mathcal{T}$ is fed to the feature extractor $G_f$ for mapping to a high-dimensional feature vector $\mathbf{f}=G_f(\mathbf{x},\theta_f)$ with parameters $\theta_f$.

The structure of feature extractor $G_f$, specifically, is shown in Fig.~\ref{feature_extractor}. It is based on a convolutional neural network (CNN) and a fully connected network (FCN) layer. The input $x_i\in\mathcal{S}\cup\mathcal{T}$ of the network is $L\times T $ images as given in Fig.~\ref{DP_configuration} (bond DP model, $T=L^z$) or in the lower part of Fig.~\ref{percolation_configuration} (2-dimensional site percolation, $T = L$). The images are convoluted by a kernel of size $3\times L$ into 16 filters forming feature maps, which indicate the locations and strength of detected features in the input. These feature maps are reduced by a factor of 2 by the max-pooling layer to reduce the size. Next, we apply a flattening layer which converts these feature maps into a vector feature $\mathbf{f}$ of size $T/2\times 16$, followed by a sigmoid activation so that the value of output is in $[0,1]$. To take this data and combine the features with a wider variety of attributes, finally, we apply a fully connected layer with 500 neurons to make the network more capable. The feature vectors in $\mathbf{f}$ are produced by applying additional batch normalization, dropout (with rate 0.8) for solving overfitting problems, and hard sigmoid map for faster and more stable results.

Those feature variables of $\mathbf{f}$ are fed into the label predictor $G_y(\mathbf{f},\theta_y)$ and domain classifier $G_d(\mathbf{f},\theta_d)$ with parameters $\theta_y$ and $\theta_d$. In detailed structure, the label predictor and domain classifier have a very similar architecture based on a fully connected layer, except the differences in output dimension and the activation function. In the label predictor we have 2 neurons and the softmax activation function, while the domain classifier has only 1 neuron with hard sigmoid activation function. Batch normalization is adopted for both before the data flow passes through the activation function. The output of label predictor is a two-dimensional vector whose two elements denote the probabilities of the configurations belonging to category ``0" (non-percolating phase, i.e., $p < p_{c}$) and category ``1" (percolating phase, i.e., $p > p_{c}$), respectively. Due to the softmax activator, the sum of elements of the vector is always 1. Additionally, the output $d$ of the domain classifier unveils the probability that the input image of the feature extractor belongs to source domain (1-$d$ is the probability of input image being taken from the target domain).

As we know, only input $x_i$ in source domain has its label as $\mathcal{S} = \{(x_s ,y_s )\}$, so the loss of the label predictor $L_y$ can only be calculated by the feature variables from source domain. On the other hand, all $x_i\in\mathcal{S}\cup\mathcal{T}$ have domain labels ($d_i=0$ for $x_i \in \mathcal{S}$ or $d_i=1$ for $x_i \in \mathcal{T}$. ), the loss of the domain classifiers $L_d$ can be addressed by full feature vector $\mathbf{f}$ of data from source and target domain.

To predict the labels of unlabeled $x_i$ in target domain, we need to ``cheat'' the domain classifier so that it can't tell which domain the feature variables come from. At the same time, the label predictor is also required to correctly distinguish labels, i.e, maximize the accuracy of the label predictor. To achieve both goals, the loss function of DANN is designed as\cite{ganin2016domain}:
\begin{equation}
	L(\theta_f, \theta_y, \theta_d)=L_y( \theta_f, \theta_y) - L_d( \theta_f, \theta_d).
\label{loss_DANN}
\end{equation}
The training process is to optimize $L$ by finding the saddle points $\hat{\theta}_f, \hat{\theta}_y$ and $\hat{\theta}_d$ with
\begin{equation}
 \begin{aligned}
	\hat{\theta}_f, \hat{\theta}_y &= \mathop{\arg\min}\limits_{\theta_f, \theta_y} L(\theta_f, \theta_y, \hat{\theta}_d),\\
	 \hat{\theta}_d &= \mathop{\arg\max}\limits_{\theta_d} L(\hat{\theta}_f, \hat{\theta}_y, \theta_d),
\end{aligned}
\label{saddle_points}
\end{equation}
which can be found as stationary points of the gradient updates
 \begin{align}
 \label{gd_f} &\theta_y \quad\leftarrow \quad\theta_f-\mu\left(\frac{\partial L_y}{\partial\theta_f}- \frac{\partial L_d}{\partial\theta_f}\right),\\ 
 \label{gd_y} &\theta_y \quad\leftarrow \quad\quad \theta_y- \mu \frac{\partial L_y}{\partial\theta_y},\\ 
 \label{gd_d} &\theta_d \quad \leftarrow \quad \quad \theta_d- \mu \frac{\partial L_d}{\partial\theta_d}.
 \end{align}
Here $\mu$ is learning rate set to 0.0001. Note, that in the domain classification loss appears with opposite signs in the gradient descent Eq.~\eqref{gd_f} for the feature extractor and the domain classifier. This trick forces DANN to do its best to classify domains and in parallel, to find features in the feature extractor, which contain the minimum possible information about domains.

This optimization process has an adversarial character and leads the feature extractor to produce a domain invariant feature vectors so that the domain classifier can't distinguish the data source. Consequently, it is also expected to reduce the effects from the different possible uneven sampling of the configurations, and please refer to~\cite{ganin2016domain} for a specific theoretical explanation of this network structure. The Adam optimizer~\cite{kingma2017adam} is used to speed up the training process of our neural network. Our adversarial domain adaptation is implemented based on TensorFlow 1.15 on AMD VEGA56 GPU platform.

\subsection{The DANN results} \label{DANN_result}
\subsubsection{Data sets of models}

In order to implement the program sketch above, one needs to prepare the datasets: the full configurations for the DP and the largest cluster configurations for the site percolation. Furthermore, we need to label part of the data. Since our goal is to minimize the human intervention, we automatically label each configuration far away from the critical regime of the phase transition as ``0" below the phase transition point (configuration dies out for DP, or, percolating cluster is missing for site percolation), and as ``1" above the phase transition point. Since for systems whose sizes are large enough, the transition is usually quite sharp, keeping "appropriate distance" from the transition point, and the number of mislabeled configurations is negligible and should be handled properly by the network through the classification. In the paper we use the criterion for the "appropriate distance" that for given probability $p$ all the configurations have in 99\% the same DANN predicted label.
Hence, for both models studied, we chose configurations generated in the $p$ range $[0,0.1] \cup [0.9,1]$ to be the initial source domain with all configurations in the range $[0,0.1]$ having label ``0" and those in the $[0.9,1]$ having label ``1". All the other configurations are considered to be in the target domain and unlabeled. Since in this paper we sampled the parameter $p$ at 41 values chosen from the range $[0,1]$ uniformly, and for each value we have generated $2000$ samples, the number of configurations in the target domain is much larger than the ones in the source domain.
For teaching we used 1000 epochs for each training set.

After training, we use the target domain configurations to predict the classification of a configuration at each value of $p$ and average them for each $p$, separately. The transition point of $p$ is found, when half of the configurations belongs to class ``0" and the other half to ``1". To obtain the critical point of a model for an infinite system, we calculate the critical points at different system sizes ($L = 16, 32, 48, 64, 80$), and extrapolate the results to an infinite system using linear regression in $1/L$.

\subsubsection{Finding the optimal source domain}
\begin{figure*}[!thb]
\subfloat[]{\includegraphics[height=45mm]{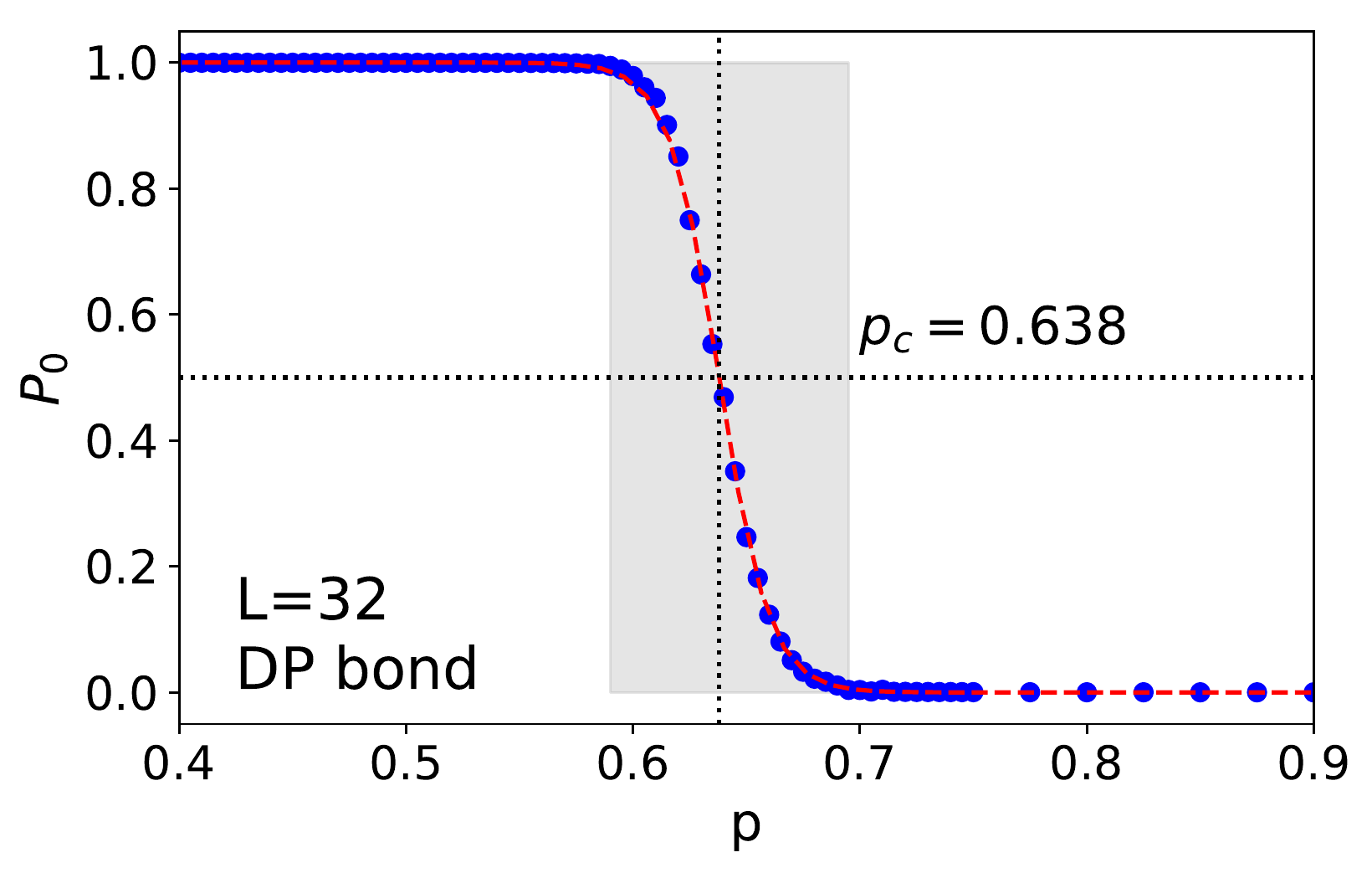}\label{DP_pc_prob}}
\subfloat[]{\includegraphics[height=45mm]{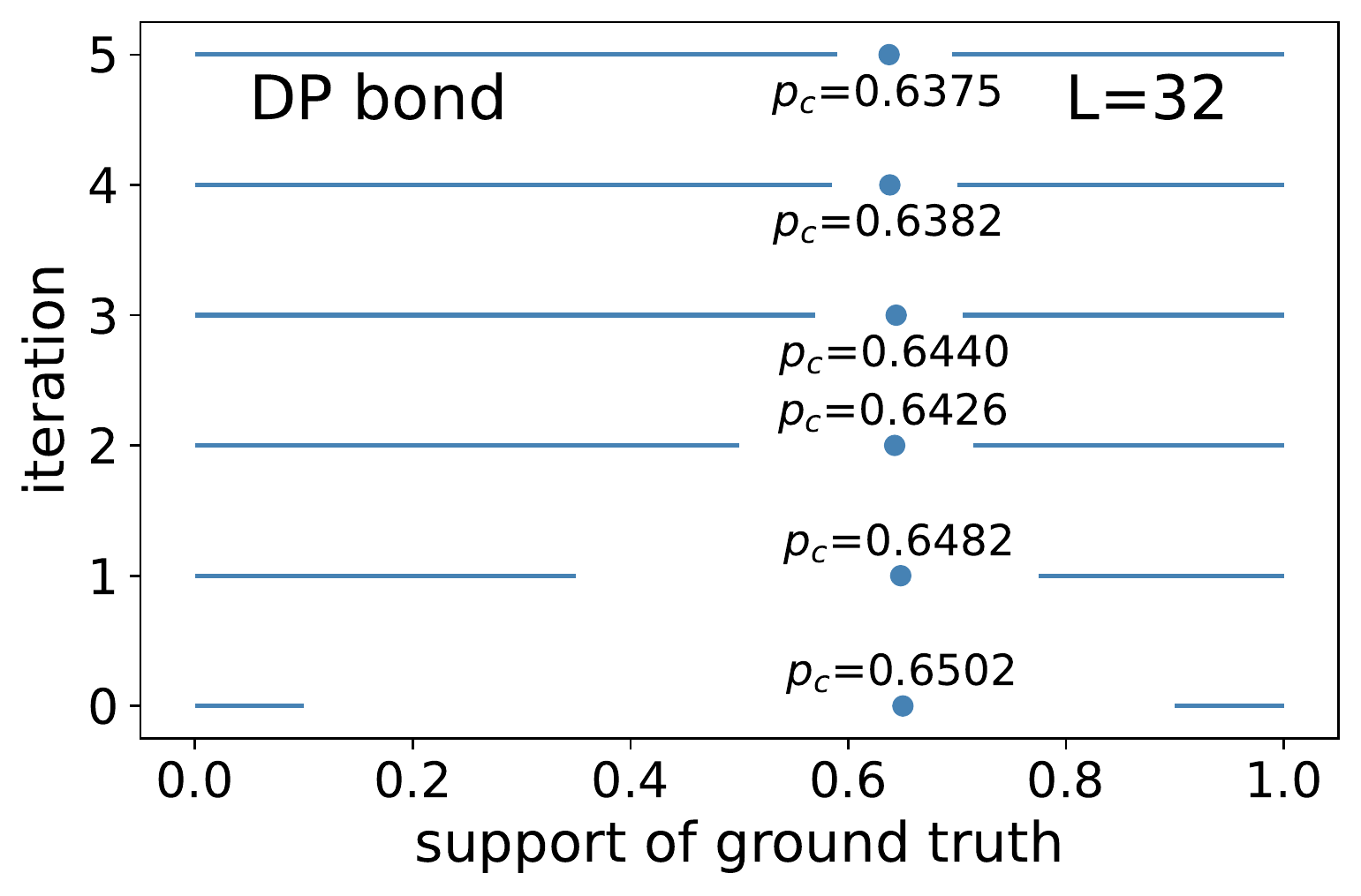}\label{DP_pc_iter}}
\caption{{\bf (a)} Average probability of belonging to phase ``0" (\tcr{$P_0$}) for DP bond percolation at $L=32$, as a function of the bond probability $p$. The shadowed region indicates the target domain at the optimal support, and the dashed red line is the sigmoid fitting to the data. Its position parameter defines the critical probability $p_c$. {\bf (b)} Evolution of the optimal domain support, and the corresponding critical bond probability values.}
\label{dp_fit_criticalpoint}
\end{figure*}

Since the chosen source domain is quite far from the expected transition point, in the following we are trying to narrow down the region of the transition, extending the original $[0,0.1] \cup [0.9,1]$ support, iteratively. We start from $[0,l] \cup [r,1]$ support with $l=0.1$ and $r=0.9$, estimating the transition probability $p_c^{0}$ with DANN, and update the domain parameters as
\begin{equation}
    l^{(i+1)} = \frac{l^{(i)}+p_c^{(i)}}{2} \,, \quad 
    r^{(i+1)} = \frac{r^{(i)}+p_c^{(i)}}{2} \,.
\label{iterate}
\end{equation}
Next, we check whether the new bounds ($l^{(i+1)}$ and $r^{(i+1)}$) fulfill the condition, based on which the datasets are categorized with at least 99\% confidence into any of the two phases. If not, the value of the bound is shifted closer to the original value, for instance $l^{(i+1),1} \to (l^{(i+1),0})/2$, and the correction is done while the confidence condition is not satisfied. The procedure is stopped, when the support can not be longer extended on the dataset.
The procedure is demonstrated in Fig.~\ref{DP_pc_iter}.

\subsubsection{Bond DP}\label{Bond_DP}

\begin{figure*}[!thb]
\subfloat[]{\includegraphics[height=38mm]{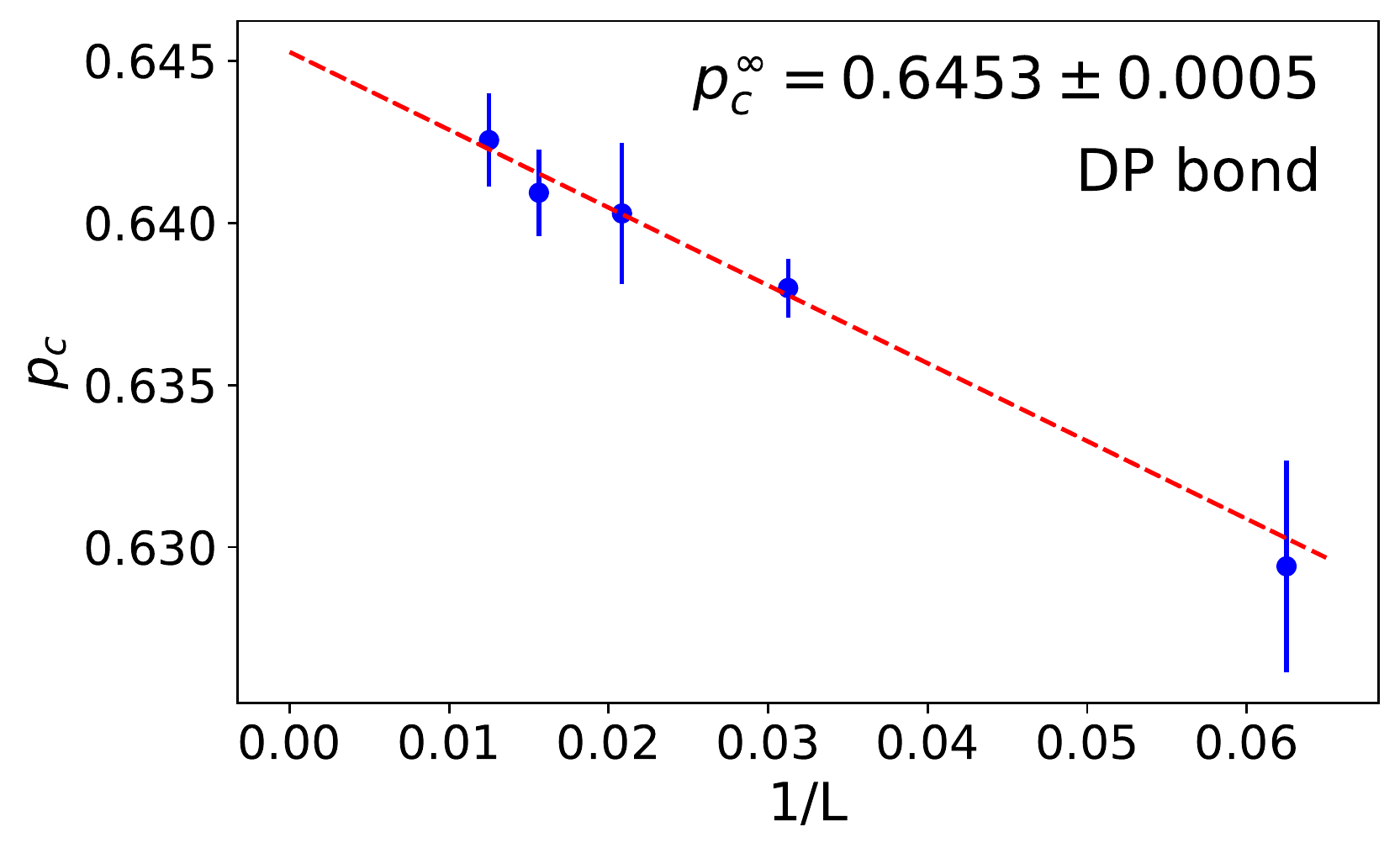}\label{DP_pc}}
\subfloat[]{\includegraphics[height=38mm]{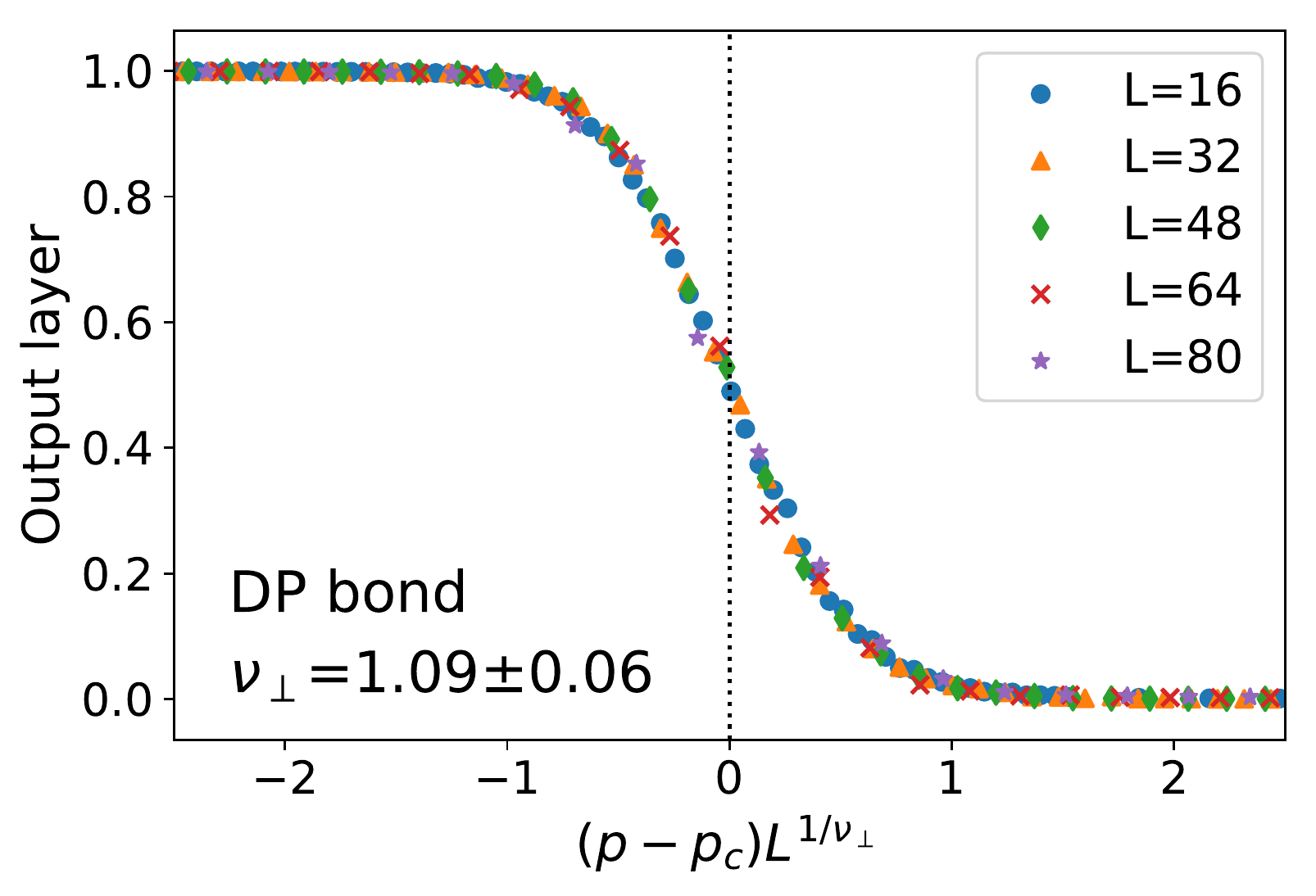}\label{DP_scale}}
\subfloat[]{\includegraphics[height=38mm]{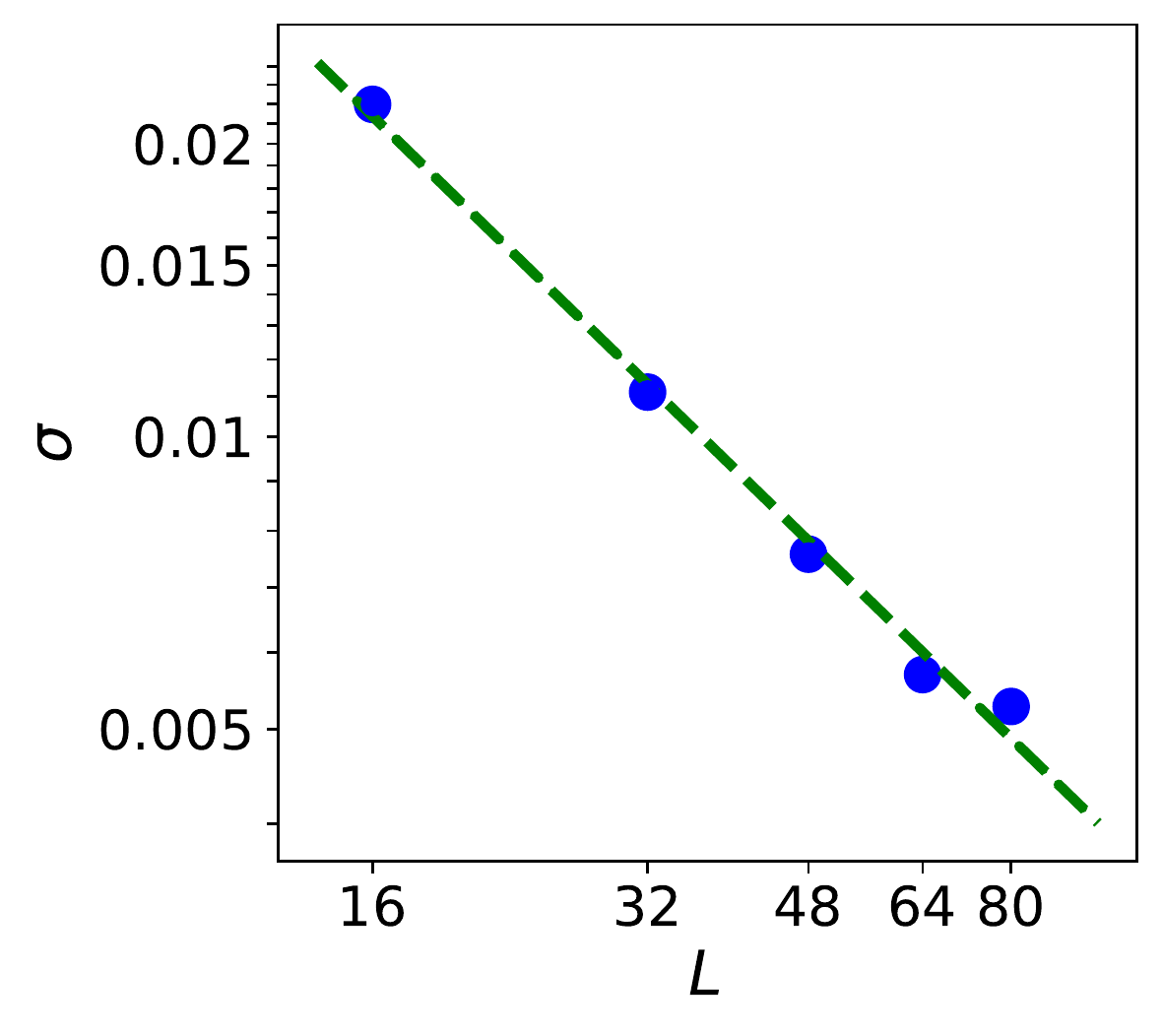}\label{DP_scale_log}}
\caption{{\bf (a)} Extrapolation of the critical probability $p_c$ to infinite lattice size for (1+1) bond DP. {\bf (b)} `Data collapse' rescaling of the results for different sizes. {\bf (c)} Fit of the critical exponent $\nu_\perp$ using the width of the sigmoid fits.}\label{DP_collapse}
\end{figure*}

After training the DANN on the optimal source domain for (1+1)-dimensional bond DP at $L = 32$, we evaluate the samples at different bond probabilities $p$, letting DANN to categorize them as either phase ``0" or phase ``1". On the output side, DANN returns a probability for each configuration's belonging to phase ``0".  The average values of the probabilities belonging to phase ``0" (\tcr{$P_0$}) are shown in Fig.~\ref{DP_pc_prob}, with a sigmoid fit. At the critical bond probability $p_c$, the probabilities of phases ``0" and ``1" are equal with each being 1/2.

Fig.~\ref{DP_pc_iter} illustrates the evolution of the optimal source domain support. The final target domain support is indicated by a grey region in Fig.~\ref{DP_pc_prob}. Note, that \tcr{only} using the very limited starting source domain ($[0,0.1] \cup [0.9,1]$) for DANN results in a reasonable estimation for the transition point.

In order to obtain the critical bond probability for the infinite system, we have trained the DANN at different sizes, and extrapolated the results to zero on the $1/L$ scale, as shown in Fig.~\ref{DP_pc}.
The acquired critical value of (1+1)-dimensional bond DP $p_{c} = 0.6453 \pm 0.0005$ is in agreement with the `standard' value of 0.6447~\cite{hinrichsen2000non}.

Using the technique of data collapse, we may obtain the value of $\nu_\perp$, the critical exponent of the spatial correlation length. The scaling $(p-p_c) L^{1/\nu_\perp}$ is universal~\cite{hinrichsen2000non}, and from our results at lattice sizes of $L = 16, 32, 48, 64$ and $80$ we may identify a proper $\nu_\perp$ to obtain the scaling (see Figs.~\ref{DP_collapse}b-c). Our value $\nu_{\perp} \simeq 1.09 \pm 0.06$ is in agreement within $1\sigma$ with the `standard' value of $1.09$~\cite{hinrichsen2000non}. We should notice that the critical exponent is not a direct result of the DANN learning, but its determination relies on the critical bond probability $p_c$ predicted by the DANN. Therefore, the accuracy of $\nu_\perp$ is related to that of $p_c$.

\subsubsection{Site Percolation}\label{Site_Percolation}

In the previous section, we have applied DANN to a typical non-equilibrium phase transition model, and found that it can find the transition point in a many-body localization problem. To check the versatility of DANN, we intend to apply it to a specific equilibrium phase transition, the two-dimensional site percolation. In site percolation, each site is occupied with probability $p$, hence the particle density depends proportionally on this probability, and is therefore not an order parameter. Hence, the full configuration, carrying information primarily on the density, can not directly reveal the order parameter.

\begin{figure*}[!htb]
\subfloat[]{\includegraphics[height=45mm]{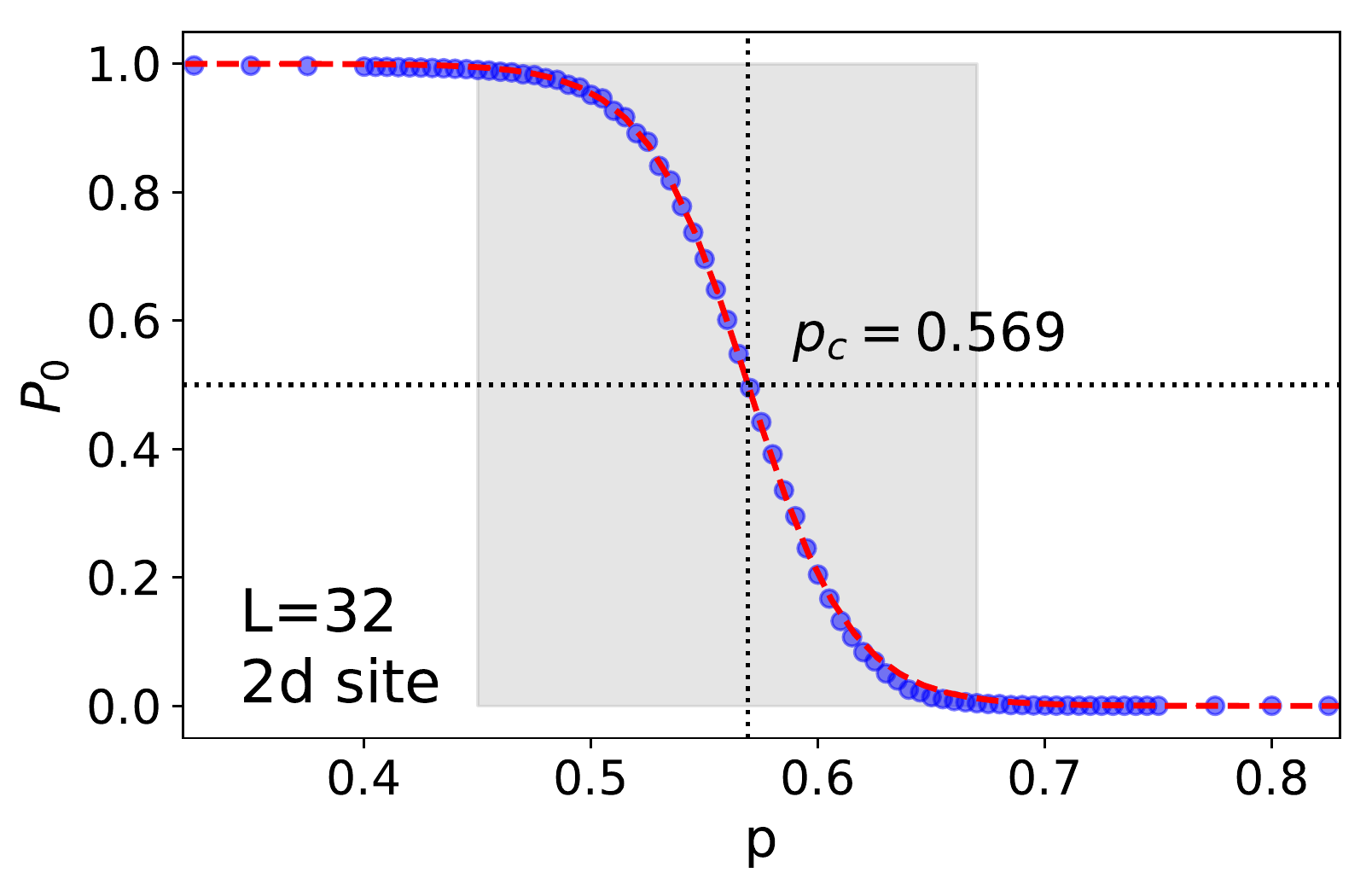}}
\subfloat[]{\includegraphics[height=45mm]{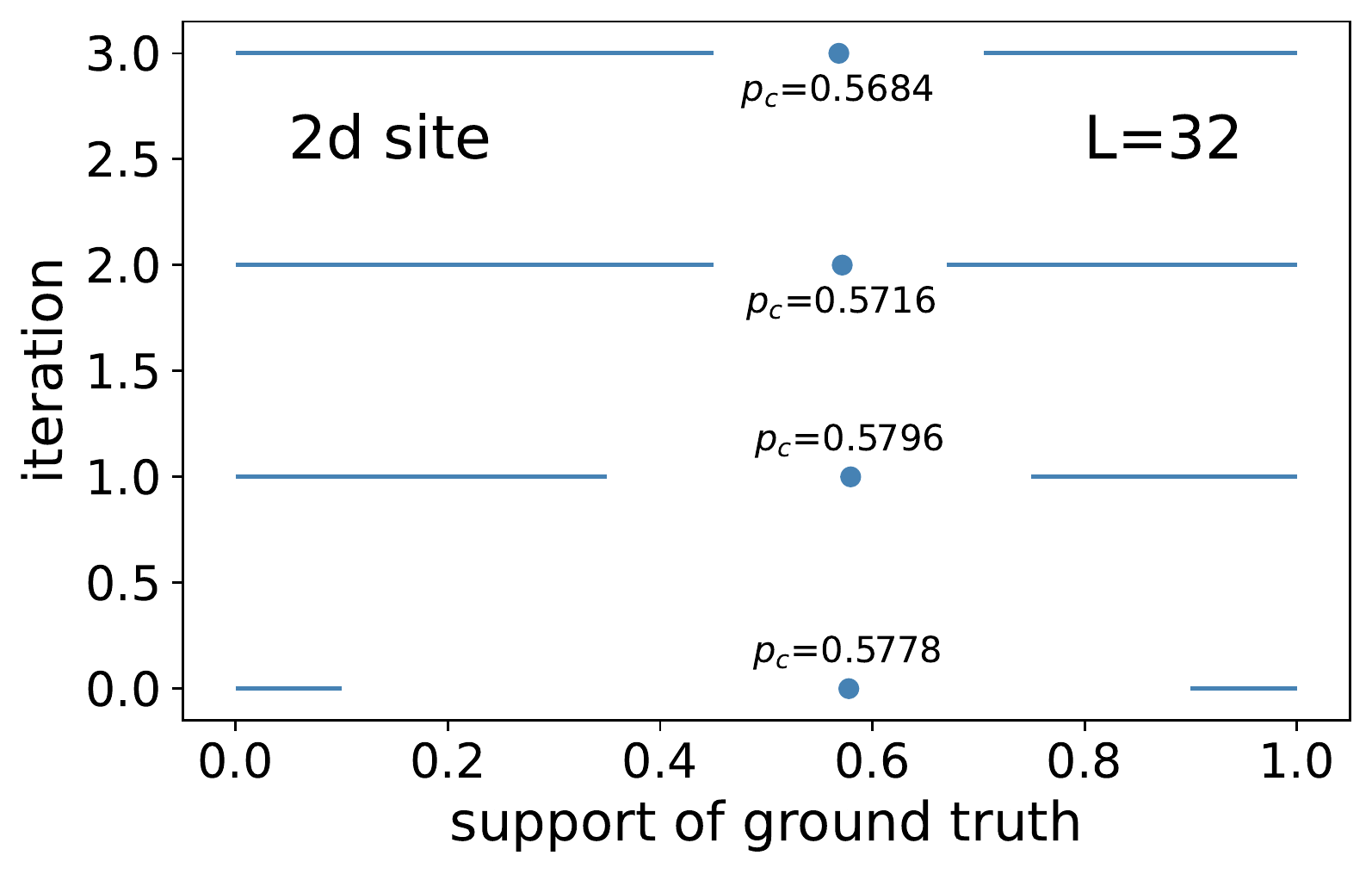}}
\caption{{\bf (a)} Average probability of belonging to phase ``0" (\tcr{$P_0$}) for two-dimensional site percolation at $L=32$, as a function of the occupation probability $p$. The shadowed region indicates the target domain at the optimal support, and the dashed line (\tcr{red in color figure}) is the sigmoid fitted to the data. Its position parameter defines the critical probability $p_c$. {\bf (b)} Evolution of the optimal domain support, and the corresponding critical bond probability values.}
\label{perc_fit_criticalpoint}
\end{figure*}

\begin{figure*}[!htb]
\subfloat[]{\includegraphics[height=38mm]{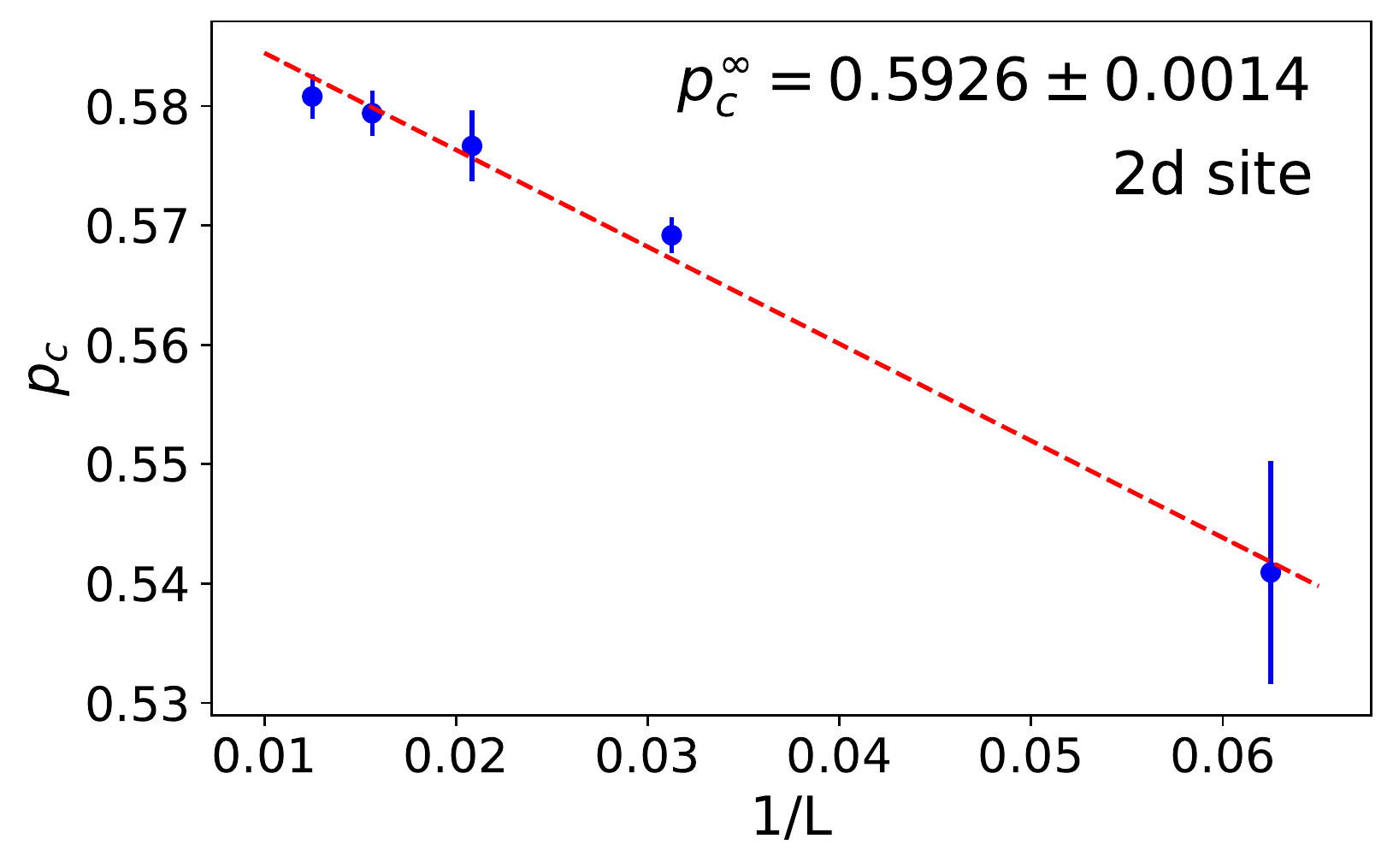}}
\subfloat[]{\includegraphics[height=38mm]{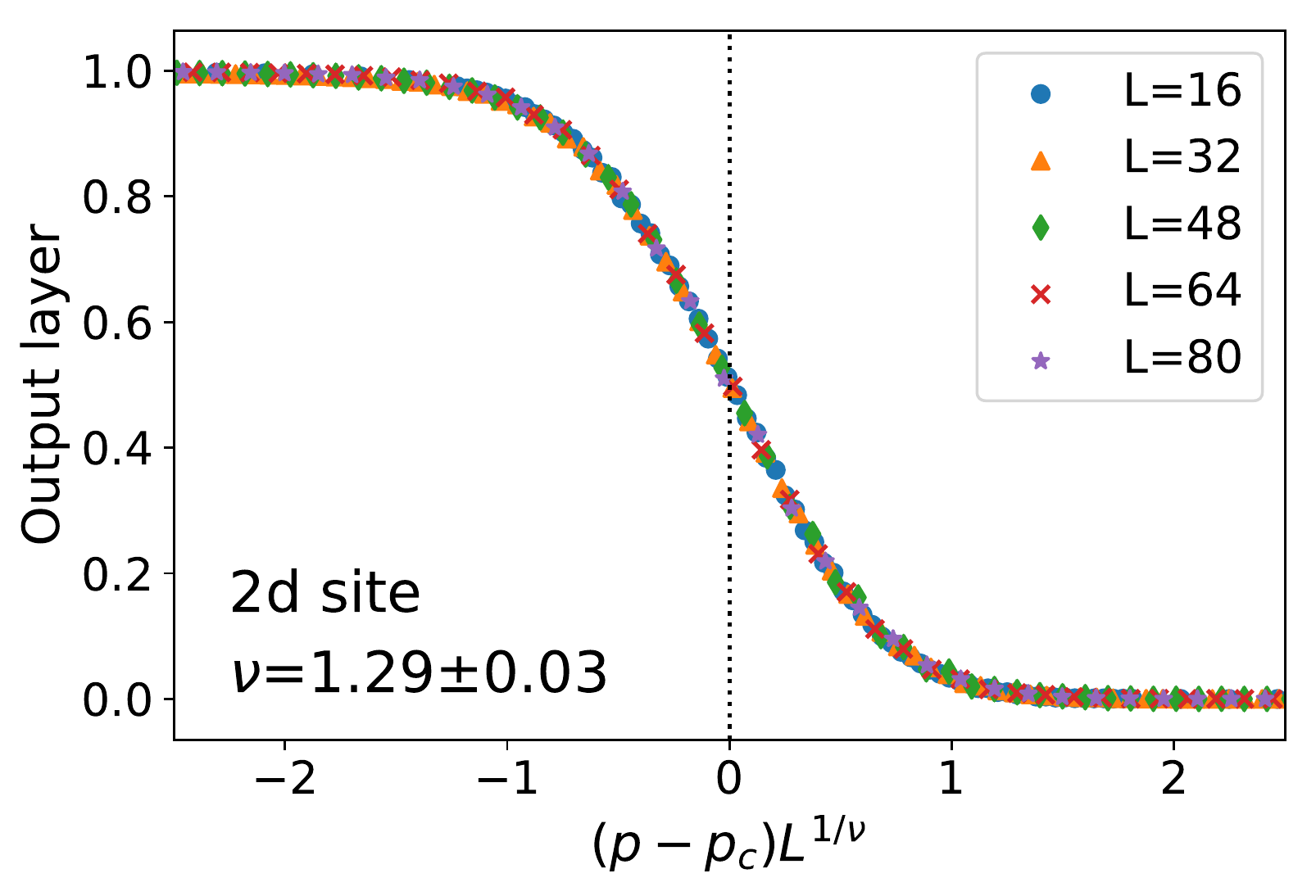}}
\subfloat[]{\includegraphics[height=38mm]{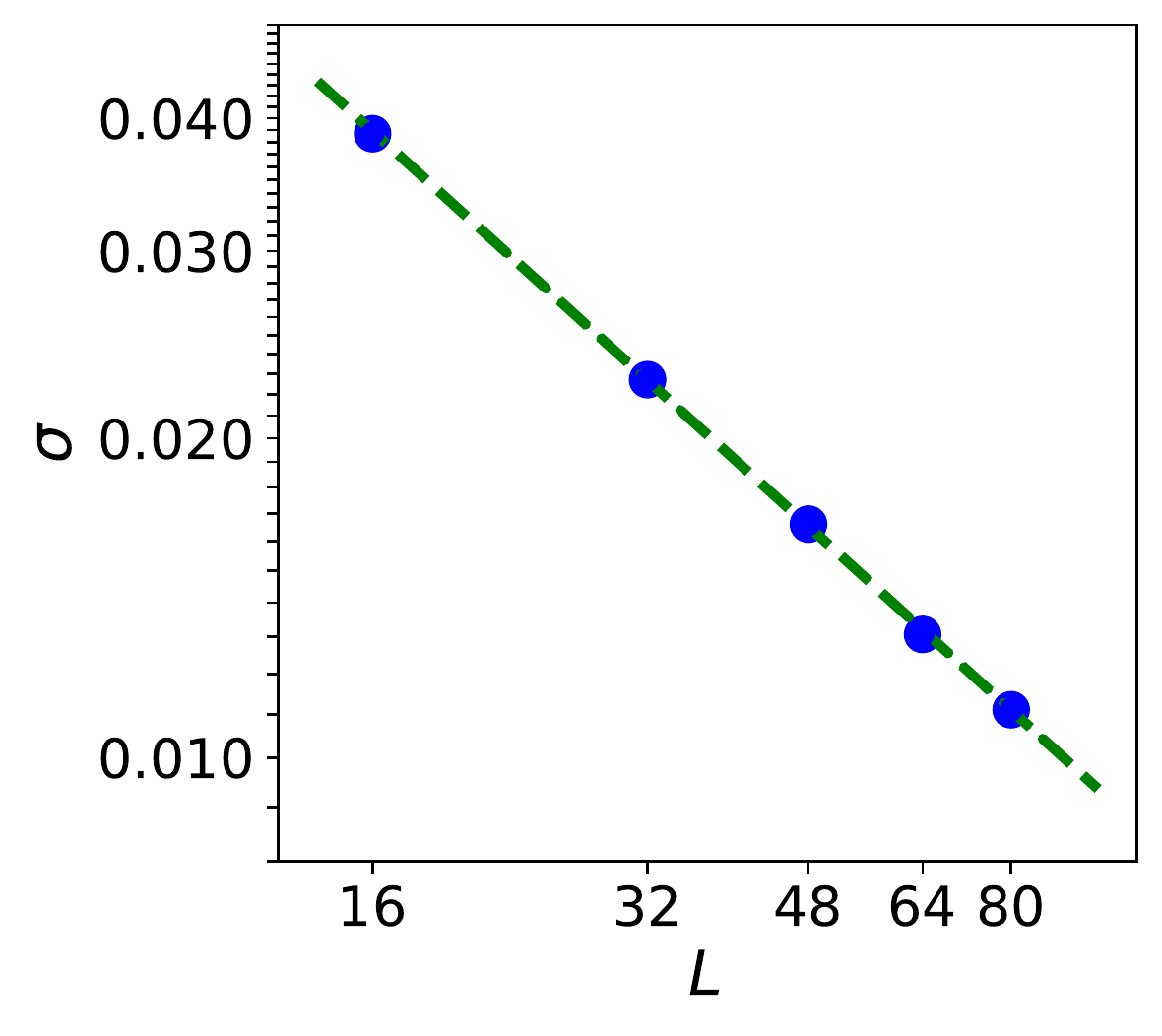}}
\caption{{\bf (a)} Extrapolation of the critical probability $p_c$ to infinite lattice size for two-dimensional site percolation. {\bf (b)} `Data collapse' rescaling of the results for different sizes. {\bf (c)} Fit of the critical exponent $\nu$ by using the width of the sigmoid fits.}
\label{perc_fit_extracriticalpoint}
\end{figure*}

The order parameter for site percolation is the probability of a site belonging to the infinite (percolating) cluster. For finite systems it can be related to the size of the largest cluster. So here we will only analyze the largest cluster: from a full configuration (upper part of Fig.~\ref{percolation_configuration}) we leave out all the sites not belonging to the largest cluster (lower part of Fig.~\ref{percolation_configuration}).
To illustrate the strong dependence of ML algorithms on the choice of data we present the learning result in Sec.~\ref{TSODPFIPT}, as a comparison, where DANN is trained with the full configurations.

We use the same DANN architecture as in the previous section and start again presenting result obtained on a $L=32$ lattice in Fig.~\ref{perc_fit_criticalpoint}. The sigmoid fit with the target domain window is presented on the left, and the determination of the optimal source domain support, on the right. Again, the procedure is repeated for sizes $L=$16, 32, 48, 64 and 80, and the critical site occupation probability is extrapolated to the infinite system in Fig.~\ref{perc_fit_extracriticalpoint}. The DANN result  $p_{c} = 0.5926 \pm 0.002$ is consistent with the numerical result $0.5927$~\cite{christensen2005complexity}, within $1\sigma$. 
The correlation exponent is obtained similarly to the DP case via data collapse, yielding $\nu \simeq 1.29 \pm 0.03$, close to $\nu = 4/3$ from~\cite{christensen2005complexity}.

\subsection{Discussions}\label{Discussions}

\subsubsection{Data preprocessing for DANN analysis}\label{TSODPFIPT}

\begin{figure*}[!thb]
 \includegraphics[height=45mm]{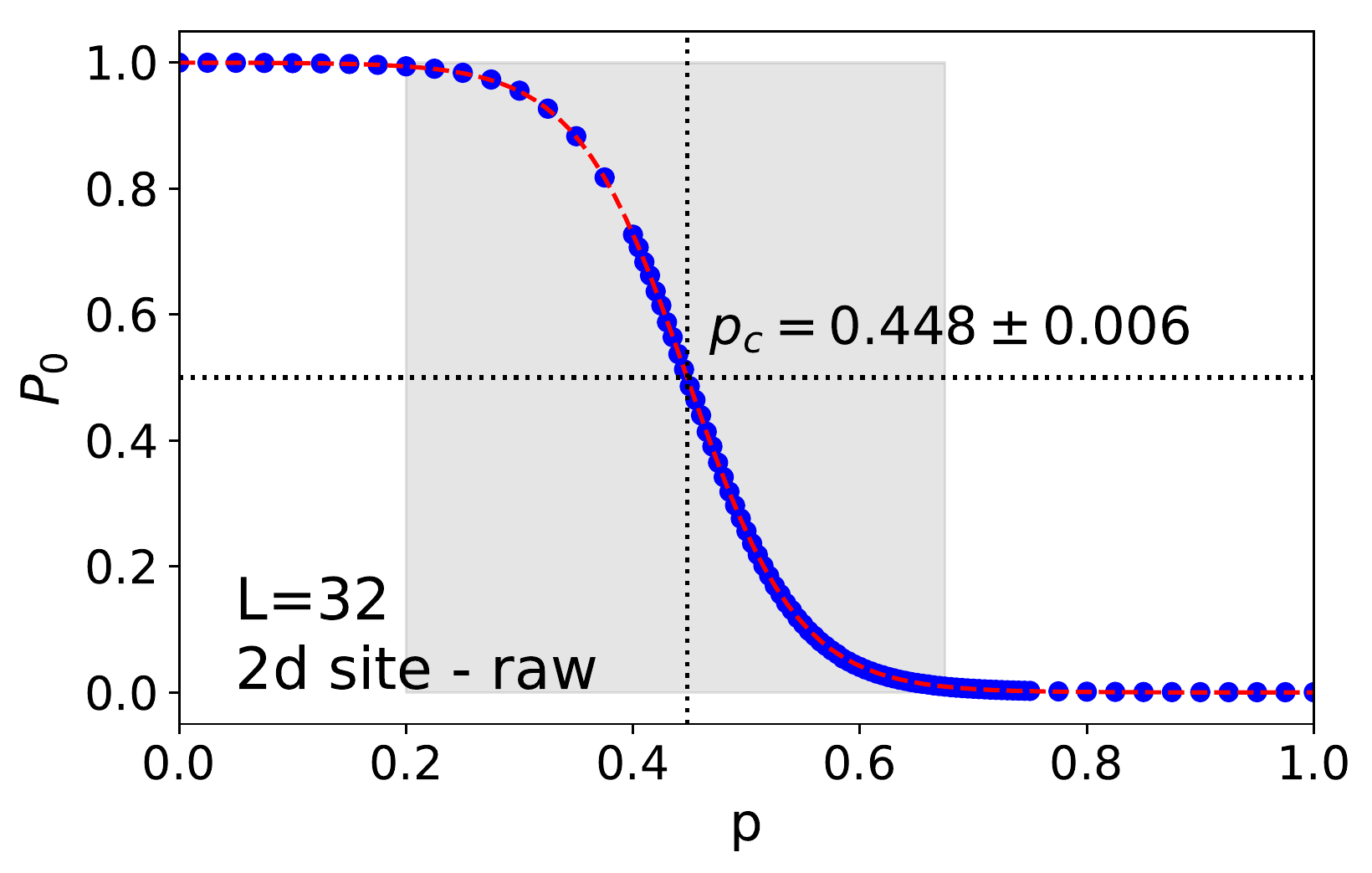}
\caption{DANN results of two-dimensional site percolation with {\em raw} configuration as input to DANN at $L = 32$. The shadowed region is the target domain at its optimal value, and is much broader than the one for the largest cluster configurations. The learned critical probability is much lower than the genuine one.}
\label{fig:perc_raw_criticalpoint}
\end{figure*}

\begin{table*}[!thb]
	\centering
	\begin{tabular}{|r|c|c|c|l|}
        \hline
        & exponents   & Literature \cite{hinrichsen2000non,christensen2005complexity} & Supervised Learning \cite{zhang2019machine,shen2021supervised} &DANN\\
        \hline      
\multirow{2}{*}{Directed percolation} & $p_{c}$ &  0.6447  & 0.6408  &  0.6453(5)   \\
 \cline{2-2} \cline{3-3} \cline{4-4}  \cline{5-5}
  &$\nu_{\perp}$ &  1.09  &   1.09(2)           &   1.09(6) \\
 \hline
\multirow{2}{*}{Site percolation} & $p_{c}$ & 0.5927 & 0.594(2) &0.5926(2)\\
 \cline{2-2} \cline{3-3} \cline{4-4}  \cline{5-5}
 &$\nu$  &  4/3 & 4/3                         & 1.29(3)  \\
 \hline
   \end{tabular}
\caption{DANN learning results of (1+1)-dimensional bond DP and two-dimensional site percolation.}
\label{ML_results}
\end{table*}
So far we have shown that DANN performs well in identifying phase transitions of the bond DP and site percolation in section~\ref{DANN_result}. However, there is extra cost in a good representation of input data that can unveil the order parameters in a direct or an indirect way. In bond DP model, the order parameter is the occupation density that directly links to raw configurations. The features, or the occupation density here, of marginal data density are accessible so that the performance of DANN is outstanding. Similar observations were made in connection to the Ising model~\cite{wetzel2017machine,PhysRevResearch.2.023266}, where the order parameter, the magnetization $M = \sum_i  s_{i}/N$, has to be included to the loss function (together with the energy) to reproduce the known results with good accuracy. Applying the principle component analysis in~\cite{wang2016discovering}, the magnetization was identified as the order parameter, implying that the method may work for more complicated systems. For site percolation Ref. ~\cite{zhang2019machine} developed a FCN and a CNN based model to detect the phase transition. Encouraged by this, using a more elaborated DL model we hope to develop in the future a network model capable of extracting the phase transition point using only raw configurations.

To illustrate the limitations of the present method (DANN), let us now try to use the raw configurations for the two-dimensional site percolation problem, instead of the largest cluster configurations. While the latter carries information about the order parameter directly, for the former it is not the case. Repeating the technique described in section~\ref{DANN_result} for the optimal source domain we gain a falsely predicted critical occupation probability $p_{c} \simeq 0.436$ at $L = 32$ (see Fig.~\ref{fig:perc_raw_criticalpoint}), which is much smaller than the one obtained from using the largest cluster configurations, or equivalently the genuine critical point. From another point of view, the DANN, and many other ML algorithms as well, have the power of identifying the transition point, if applicable, embedded in the configurations. But the transition point may not be the actual critical point governed by the order parameter. Further information regarding the order parameter has to be compiled, which is then transferred to the learning algorithm.

\subsubsection{Comparison of the DANN results with supervised learning}\label{DANN and supervised learning}

\begin{figure*}[!thb]
\subfloat[]{\includegraphics[height=45mm]{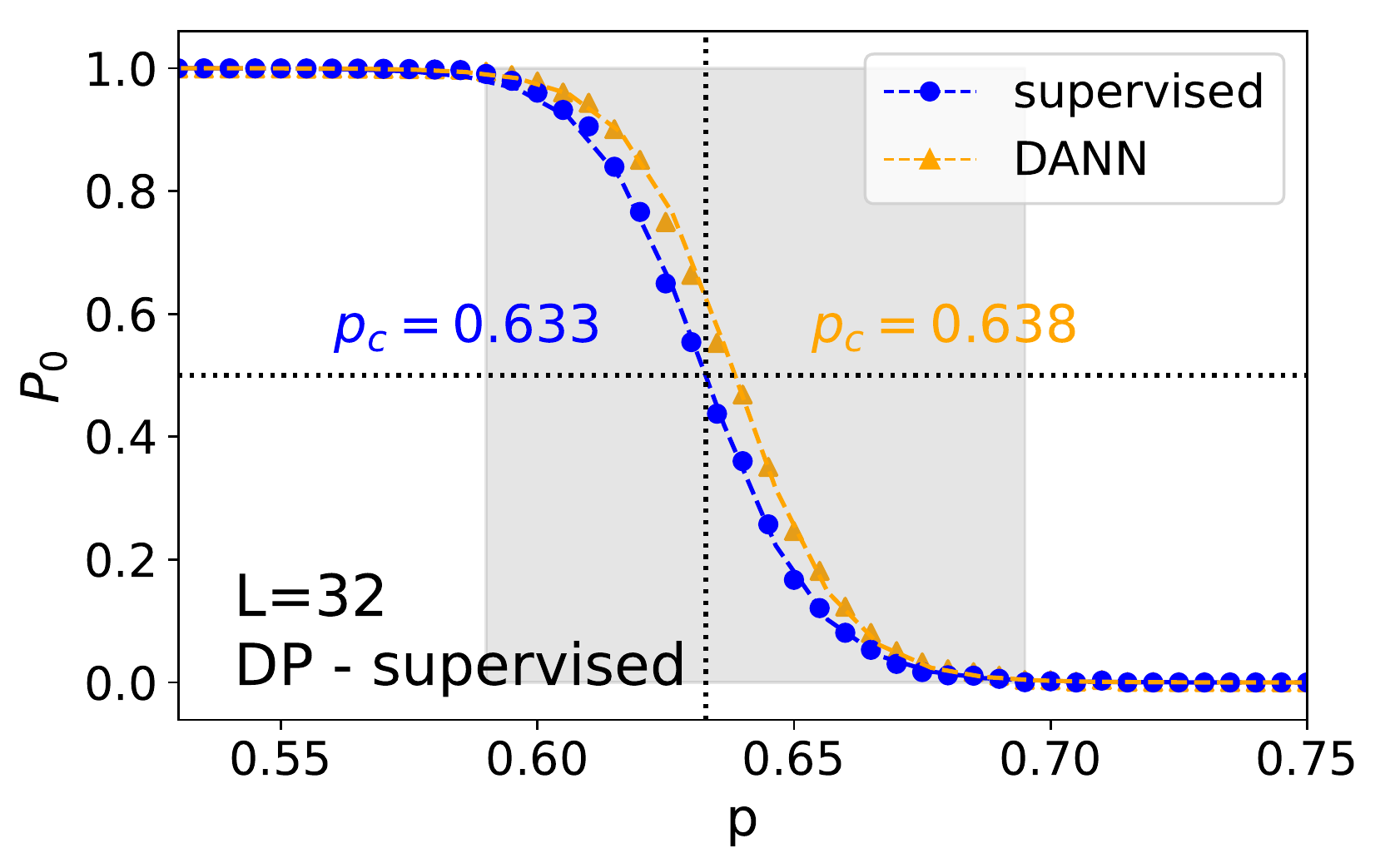}\label{DP_supervised}}
\subfloat[]{\includegraphics[height=45mm]{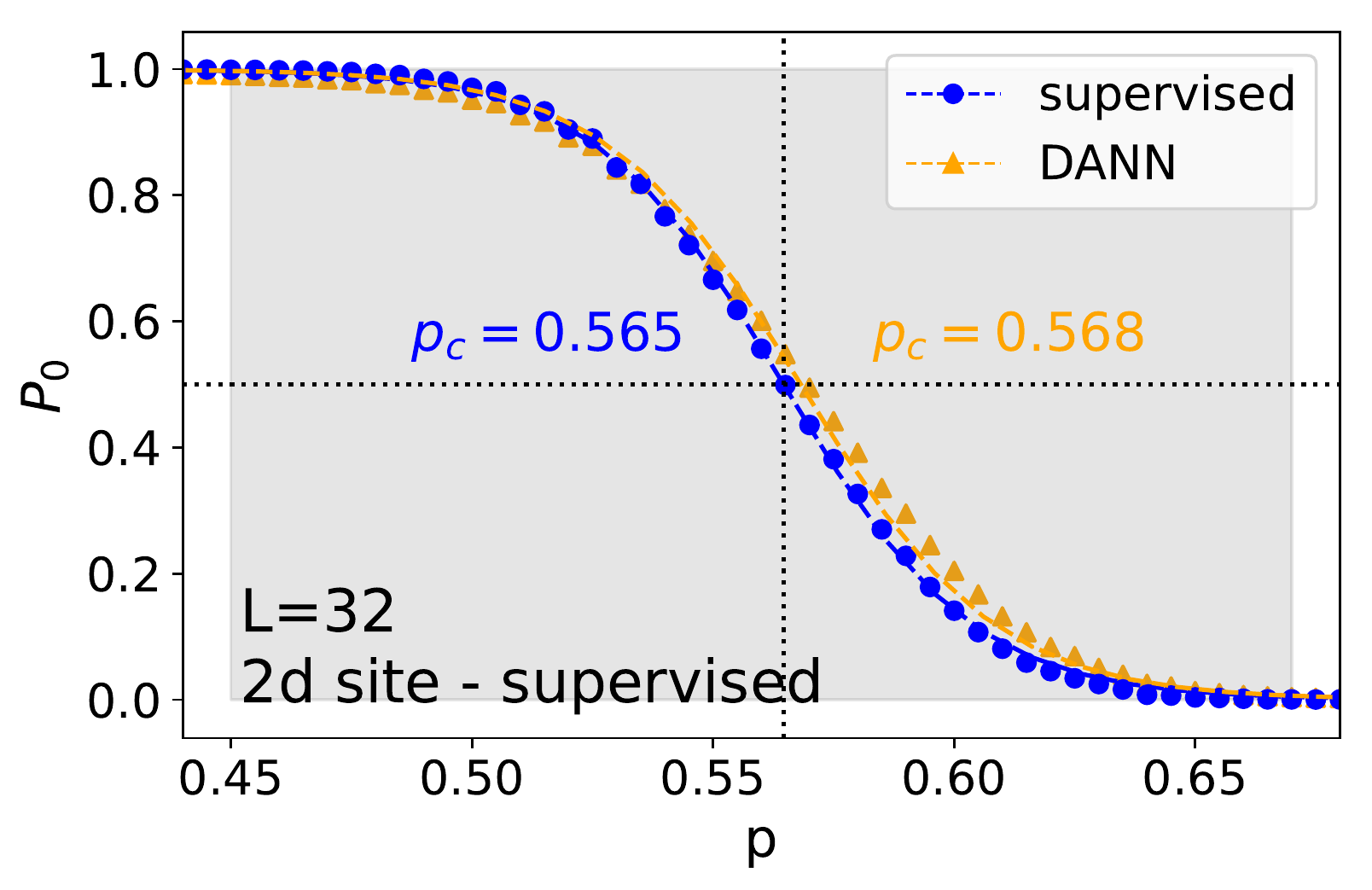}\label{perc2d_supervised}}
\caption{Supervised learning versus DANN learning of {\bf (a)} (1+1) bond DP and {\bf (b)} two-dimensional site percolation models at size $L=32$.}
\label{dp_NNandDANN_32_criticalpoint}
\end{figure*}

The models of interest (the directed (1+1) dimensional bond percolation and the two-dimensional site percolation of square lattice) have already been extensively investigated by different theoretical models (like e.g. mean-field, renormalization group), numerical Monte-Carlo simulations~\cite{hinrichsen2000non,christensen2005complexity}, or, supervised learning~\cite{zhang2019machine,shen2021supervised}. We compared major previous studies with the present study in Table~\ref{ML_results}.

The common measurement between reference \cite{zhang2019machine} of using supervised learning and our work of using \tcr{DANNs} is concerned the site percolation. We compared the runtime of the two methods with different system sizes (Please refer to Tab. \ref{tablecompucost_percolation}). We find that if the number of iterations of the optimal domain support is smaller than 8, the DANNs will be slightly time saving than supervised learning. For the critical value $p_c$, reference \cite{zhang2019machine} predicts 0.594(2) and DANNs, 0.5926(2). Both predictions of $p_c$ are close to 0.5927, the theoretical value and DANNs is slightly more favored. For the correlation exponent $\nu$, reference \cite{zhang2019machine} predicts 4/3 and DANNs, 1.29(3). The theoretical value of $\nu$ is 4/3, which favors reference \cite{zhang2019machine}. So far, neither of the two methods is able to handle larger systems size.

\begin{table*}[!tbp]
	\centering
	\begin{tabular}{|c|c|c|c|c|c|c|}
        \hline
 lattice size &  $8$ & $16$  & $32$   & $48$ & $64$ \\
        \hline
 time cost(Supervised) &  $57.3s$  & $103.9s$   &$ 301.7s$ & $681.5s$  & $1193.6s$   \\

        \hline
 time cost(DANN) & $12.9s \times i$ & $18.0s \times i$   & $42.6s \times i$ & $80.9s \times i$ & $147.9s \times i$  \\
        \hline
    \end{tabular}

\caption{The comparison of computational time (in terms of seconds) between supervised learning and semi-supervised learning (DANN) in site percolation. $i$ indicates the number of iterations in the evolution of the optimal domain support for DANN, \tcr{and is taken to be 3 for site percolation.}}
\label{tablecompucost_percolation}
\end{table*}

The common measurement of reference \cite{shen2021supervised} and DANN is concerned \tcr{ with the studying of} (1+1)-DP model. For the critical value $p_c$, reference \cite{shen2021supervised} gives 0.6408 (from supervised learning) and 0.656(1) (from unsupervised learning) and DANN of this work, 0.6453(5). The theoretical value of $p_c$ is 0.6447, which means that DANN and unsupervised learning work are nearly equally efficient in predicting the $p_c$ value. For supervised learning, the value is of not much reference as the theoretical value is an input in labelling. For the spatial correlation exponent $\nu_{\perp}$, supervised learning of reference \cite{shen2021supervised} predicts 1.09(2) and DANN, 1.09(6). Both predictions are close to the theoretical value 1.09. Tab. \ref{tablecompucost_dp} shows the runtime comparison between reference \cite{shen2021supervised} and DANN in dealing with (1+1)-DP model. As is seen, if the number of iterations of the optimal domain support is smaller than 8, the DANNs will be slightly time saving than supervised learning. Also, if the number of iterations of the optimal domain support is smaller than 2, the DANNs will be slightly time saving than unsupervised learning, namely autoencoder here.

\begin{table*}[!tbp]
	\centering
	\begin{tabular}{|c|c|c|c|c|c|c|}
        \hline
 lattice size &  $8$ & $16$  & $32$   & $48$ & $64$ \\
        \hline
 time cost(Supervised) &  $102.8s$  & $122.9s$   &$ 298.5s$ &$723.5s$  &$1332.9s$   \\

        \hline
 time cost(autoencoder) & $32.5s$ & $43.7s$   & $51.2s$ & $79.3s$ & $138.2s$  \\
        \hline
         time cost(DANN) & $13.1s \times i$ & $18.6s \times i$   & $39.9s \times i$ & $75.2s \times i$ & $127.5s \times i$  \\
        \hline
    \end{tabular}

\caption{The comparison of computational time (in terms of seconds) between supervised learning, unsupervised learning (autoencoder) and semi-supervised learning (DANN) in (1+1) dimensional DP. $i$ indicates the number of iterations in the evolution of the optimal domain support for DANN, \tcr{and is taken to be 5 for DP in our simulations.}}
\label{tablecompucost_dp}
\end{table*}

From the comparison we conclude that for this two models DANN performs nearly as well as the supervised learning does. Clearly the latter requires extra work as one has to label the different configurations.

To check the power of DANN in exploiting unknown regions, we perform the following test: we build a network similar to the DANN, leaving out the classification part (that is, dropping the adversarial character), and train the resulting networks on the optimal support of the source domain, calculated for the DANN. On the left support (source domain $[0,0.1] \cap [0.9, 1]$) we mark each configuration as belonging to phase ``0", on the right to phase ``1". The results for $L=32$ are presented in Fig.~\ref{dp_NNandDANN_32_criticalpoint}, where the shadowed region indicates the target domain (i.e. the domain without labels which are not presented to the non-adversarial network during training), showing a small, but noticeable shift in the transition point from the DANN values: $0.633$ instead of 0.638 for DP, and 0.565 instead of 0.568 for two-dimensional site percolation. The difference is much more pronounced for the starting support of the source domain, $[0,0.1] \cup [0.9,1]$ ($0.6152$ instead of $0.6502$ for DP, and $0.5975$ instead of $0.5778$ for two-dimensional site percolation), indicating that despite using a very limited dataset, the DANN is able to produce more reliable result than a non-adversarial network is.

\section{Conclusion}\label{Conclusion}

In this paper, we have applied a semi-supervised or transfer learning approach, the domain adversarial neural network (DANN), to detecting the critical points of phase transitions. Specifically, to explore the applicability of the DANN method, we have studied two kinds of models, the (1+1)-dimensional bond directed percolation as a representative of non-equilibrium systems and the two-dimensional site percolation as one of equilibrium systems. For (1+1)-dimensional bond DP model, we demonstrate how one may find the critical point $p_{c}$ and calculate the critical exponent $\nu_{\perp}$. Introducing an iterative approach to extend the source domain of the DANN, we are able to reproduce with high precision the genuine results for the transition point and critical exponent, using much smaller set of configurations than the Monte-Carlo methods.

In two-dimensional site percolation, we have shown that a careful preprocessing of the input data related to the order parameter should be applied to speeding up the training and improving the accuracy. Specifically, 
the largest cluster is singled out from the full configuration as input. Our conclusion is that DANN is a powerful method in determining the location of the critical point, and the spatial correlation exponent as well, with good accuracy and minimal computational cost. The method is then combined to the data collapse techniques, in order to have finite-size effects and then yield the values corresponding to the infinite system.

We also noted that in the present form the DANN is very sensitive to the input configuration, as it expects configurations directly related to the order parameter. Presenting the network with raw configurations of two-dimensional site percolation the accuracy of the result is dropped tremendously. This issue will be systematically studied in the near future, by enabling the network to acquire automatically the relevant parts of the configuration, hopefully also in systems where the order parameter is more explicit.

The biggest advantage of DANN is its predictive power. Supervised learning assumes that the phase transition point is already known, and then predicts the accurate location by learning the information of this known range. DANN is able to predict quite accurately the location of the transition point already from very limited labelled information: a non-adversarial network using the same information can only obtain a much less precise result. On the other hand, a non-adversarial network learns more easily the relevant part of the configuration (`pre-filtering'), since it is presented with much more labelled data.

%\FloatBarrier
\section{Acknowledgements}
We would like to thank Zhencheng Fu and Hongwei Tan for their helpful suggestions at the beginning of this work. This work was supported in part by the Fundamental Research Funds for the Central Universities, China (Grant No. CCNU19QN029), the National Natural Science Foundation of China (Grant No. 11505071, 61702207 and 61873104), the 111 Project, with Grant No. BP0820038, the Hungarian Research Fund (OTKA) under the grant number K123815 and the Ministry of Innovation and Technology NRDI Office within the framework of the MILAB Artificial Intelligence National Laboratory Program.

%\tcb{titles are not broken in the bibliography due to the {\bf ulem} package. In the final version {\bf ulem} should be removed!}

\bibliographystyle{apsrev4-2}
\bibliography{transfer_learningref}

%apsrev4-2.bst 2019-01-14 (MD) hand-edited version of apsrev4-1.bst
%Control: key (0)
%Control: author (72) initials jnrlst
%Control: editor formatted (1) identically to author
%Control: production of article title (-1) disabled
%Control: page (0) single
%Control: year (1) truncated
%Control: production of eprint (0) enabled
\begin{thebibliography}{48}%
\makeatletter
\providecommand \@ifxundefined [1]{%
 \@ifx{#1\undefined}
}%
\providecommand \@ifnum [1]{%
 \ifnum #1\expandafter \@firstoftwo
 \else \expandafter \@secondoftwo
 \fi
}%
\providecommand \@ifx [1]{%
 \ifx #1\expandafter \@firstoftwo
 \else \expandafter \@secondoftwo
 \fi
}%
\providecommand \natexlab [1]{#1}%
\providecommand \enquote  [1]{``#1''}%
\providecommand \bibnamefont  [1]{#1}%
\providecommand \bibfnamefont [1]{#1}%
\providecommand \citenamefont [1]{#1}%
\providecommand \href@noop [0]{\@secondoftwo}%
\providecommand \href [0]{\begingroup \@sanitize@url \@href}%
\providecommand \@href[1]{\@@startlink{#1}\@@href}%
\providecommand \@@href[1]{\endgroup#1\@@endlink}%
\providecommand \@sanitize@url [0]{\catcode `\\12\catcode `\$12\catcode
  `\&12\catcode `\#12\catcode `\^12\catcode `\_12\catcode `\%12\relax}%
\providecommand \@@startlink[1]{}%
\providecommand \@@endlink[0]{}%
\providecommand \url  [0]{\begingroup\@sanitize@url \@url }%
\providecommand \@url [1]{\endgroup\@href {#1}{\urlprefix }}%
\providecommand \urlprefix  [0]{URL }%
\providecommand \Eprint [0]{\href }%
\providecommand \doibase [0]{https://doi.org/}%
\providecommand \selectlanguage [0]{\@gobble}%
\providecommand \bibinfo  [0]{\@secondoftwo}%
\providecommand \bibfield  [0]{\@secondoftwo}%
\providecommand \translation [1]{[#1]}%
\providecommand \BibitemOpen [0]{}%
\providecommand \bibitemStop [0]{}%
\providecommand \bibitemNoStop [0]{.\EOS\space}%
\providecommand \EOS [0]{\spacefactor3000\relax}%
\providecommand \BibitemShut  [1]{\csname bibitem#1\endcsname}%
\let\auto@bib@innerbib\@empty
%</preamble>
\bibitem [{\citenamefont {Jordan}\ and\ \citenamefont
  {Mitchell}(2015)}]{jordan2015machine}%
  \BibitemOpen
  \bibfield  {author} {\bibinfo {author} {\bibfnamefont {M.~I.}\ \bibnamefont
  {Jordan}}\ and\ \bibinfo {author} {\bibfnamefont {T.~M.}\ \bibnamefont
  {Mitchell}},\ }\href@noop {} {\bibfield  {journal} {\bibinfo  {journal}
  {Science}\ }\textbf {\bibinfo {volume} {349}},\ \bibinfo {pages} {255}
  (\bibinfo {year} {2015})}\BibitemShut {NoStop}%
\bibitem [{\citenamefont {Goodfellow}\ \emph {et~al.}(2016)\citenamefont
  {Goodfellow}, \citenamefont {Bengio},\ and\ \citenamefont
  {Courville}}]{goodfellow2016machine}%
  \BibitemOpen
  \bibfield  {author} {\bibinfo {author} {\bibfnamefont {I.}~\bibnamefont
  {Goodfellow}}, \bibinfo {author} {\bibfnamefont {Y.}~\bibnamefont {Bengio}},\
  and\ \bibinfo {author} {\bibfnamefont {A.}~\bibnamefont {Courville}},\
  }\href@noop {} {\bibfield  {journal} {\bibinfo  {journal} {Deep learning}\
  }\textbf {\bibinfo {volume} {1}},\ \bibinfo {pages} {98} (\bibinfo {year}
  {2016})}\BibitemShut {NoStop}%
\bibitem [{\citenamefont {Engel}\ and\ \citenamefont {Van~den
  Broeck}(2001)}]{engel2001statistical}%
  \BibitemOpen
  \bibfield  {author} {\bibinfo {author} {\bibfnamefont {A.}~\bibnamefont
  {Engel}}\ and\ \bibinfo {author} {\bibfnamefont {C.}~\bibnamefont {Van~den
  Broeck}},\ }\href@noop {} {\emph {\bibinfo {title} {Statistical mechanics of
  learning}}}\ (\bibinfo  {publisher} {Cambridge University Press},\ \bibinfo
  {year} {2001})\BibitemShut {NoStop}%
\bibitem [{\citenamefont {Mehta}\ and\ \citenamefont
  {Schwab}(2014)}]{mehta2014exact}%
  \BibitemOpen
  \bibfield  {author} {\bibinfo {author} {\bibfnamefont {P.}~\bibnamefont
  {Mehta}}\ and\ \bibinfo {author} {\bibfnamefont {D.~J.}\ \bibnamefont
  {Schwab}},\ }\href@noop {} {\bibfield  {journal} {\bibinfo  {journal} {arXiv
  preprint arXiv:1410.3831}\ } (\bibinfo {year} {2014})}\BibitemShut {NoStop}%
\bibitem [{\citenamefont {Mehta}\ \emph {et~al.}(2019)\citenamefont {Mehta},
  \citenamefont {Bukov}, \citenamefont {Wang}, \citenamefont {Day},
  \citenamefont {Richardson}, \citenamefont {Fisher},\ and\ \citenamefont
  {Schwab}}]{mehta2019high}%
  \BibitemOpen
  \bibfield  {author} {\bibinfo {author} {\bibfnamefont {P.}~\bibnamefont
  {Mehta}}, \bibinfo {author} {\bibfnamefont {M.}~\bibnamefont {Bukov}},
  \bibinfo {author} {\bibfnamefont {C.-H.}\ \bibnamefont {Wang}}, \bibinfo
  {author} {\bibfnamefont {A.~G.}\ \bibnamefont {Day}}, \bibinfo {author}
  {\bibfnamefont {C.}~\bibnamefont {Richardson}}, \bibinfo {author}
  {\bibfnamefont {C.~K.}\ \bibnamefont {Fisher}},\ and\ \bibinfo {author}
  {\bibfnamefont {D.~J.}\ \bibnamefont {Schwab}},\ }\href@noop {} {\bibfield
  {journal} {\bibinfo  {journal} {Physics reports}\ }\textbf {\bibinfo {volume}
  {810}},\ \bibinfo {pages} {1} (\bibinfo {year} {2019})}\BibitemShut {NoStop}%
\bibitem [{\citenamefont {Carleo}\ \emph {et~al.}(2019)\citenamefont {Carleo},
  \citenamefont {Cirac}, \citenamefont {Cranmer}, \citenamefont {Daudet},
  \citenamefont {Schuld}, \citenamefont {Tishby}, \citenamefont
  {Vogt-Maranto},\ and\ \citenamefont {Zdeborov{\'a}}}]{carleo2019machine}%
  \BibitemOpen
  \bibfield  {author} {\bibinfo {author} {\bibfnamefont {G.}~\bibnamefont
  {Carleo}}, \bibinfo {author} {\bibfnamefont {I.}~\bibnamefont {Cirac}},
  \bibinfo {author} {\bibfnamefont {K.}~\bibnamefont {Cranmer}}, \bibinfo
  {author} {\bibfnamefont {L.}~\bibnamefont {Daudet}}, \bibinfo {author}
  {\bibfnamefont {M.}~\bibnamefont {Schuld}}, \bibinfo {author} {\bibfnamefont
  {N.}~\bibnamefont {Tishby}}, \bibinfo {author} {\bibfnamefont
  {L.}~\bibnamefont {Vogt-Maranto}},\ and\ \bibinfo {author} {\bibfnamefont
  {L.}~\bibnamefont {Zdeborov{\'a}}},\ }\href@noop {} {\bibfield  {journal}
  {\bibinfo  {journal} {Reviews of Modern Physics}\ }\textbf {\bibinfo {volume}
  {91}},\ \bibinfo {pages} {045002} (\bibinfo {year} {2019})}\BibitemShut
  {NoStop}%
\bibitem [{\citenamefont {Carrasquilla}(2020)}]{carrasquilla2020machine}%
  \BibitemOpen
  \bibfield  {author} {\bibinfo {author} {\bibfnamefont {J.}~\bibnamefont
  {Carrasquilla}},\ }\href@noop {} {\bibfield  {journal} {\bibinfo  {journal}
  {Advances in Physics: X}\ }\textbf {\bibinfo {volume} {5}},\ \bibinfo {pages}
  {1797528} (\bibinfo {year} {2020})}\BibitemShut {NoStop}%
\bibitem [{\citenamefont {Domb}(1996)}]{domb1996critical}%
  \BibitemOpen
  \bibfield  {author} {\bibinfo {author} {\bibfnamefont {C.}~\bibnamefont
  {Domb}},\ }\href@noop {} {\emph {\bibinfo {title} {The critical point: a
  historical introduction to the modern theory of critical phenomena}}}\
  (\bibinfo  {publisher} {CRC Press},\ \bibinfo {year} {1996})\BibitemShut
  {NoStop}%
\bibitem [{\citenamefont {Amit}\ and\ \citenamefont
  {Martin-Mayor}(2005)}]{amit2005field}%
  \BibitemOpen
  \bibfield  {author} {\bibinfo {author} {\bibfnamefont {D.~J.}\ \bibnamefont
  {Amit}}\ and\ \bibinfo {author} {\bibfnamefont {V.}~\bibnamefont
  {Martin-Mayor}},\ }\href@noop {} {\emph {\bibinfo {title} {Field Theory, the
  Renormalization Group, and Critical Phenomena: Graphs to Computers Third
  Edition}}}\ (\bibinfo  {publisher} {World Scientific Publishing Company},\
  \bibinfo {year} {2005})\BibitemShut {NoStop}%
\bibitem [{\citenamefont {Hammersley}(2013)}]{hammersley2013monte}%
  \BibitemOpen
  \bibfield  {author} {\bibinfo {author} {\bibfnamefont {J.}~\bibnamefont
  {Hammersley}},\ }\href@noop {} {\emph {\bibinfo {title} {Monte carlo
  methods}}}\ (\bibinfo  {publisher} {Springer Science \& Business Media},\
  \bibinfo {year} {2013})\BibitemShut {NoStop}%
\bibitem [{\citenamefont {Carrasquilla}\ and\ \citenamefont
  {Melko}(2017)}]{carrasquilla2017machine}%
  \BibitemOpen
  \bibfield  {author} {\bibinfo {author} {\bibfnamefont {J.}~\bibnamefont
  {Carrasquilla}}\ and\ \bibinfo {author} {\bibfnamefont {R.~G.}\ \bibnamefont
  {Melko}},\ }\href@noop {} {\bibfield  {journal} {\bibinfo  {journal} {Nature
  Physics}\ }\textbf {\bibinfo {volume} {13}},\ \bibinfo {pages} {431}
  (\bibinfo {year} {2017})}\BibitemShut {NoStop}%
\bibitem [{\citenamefont {Van~Nieuwenburg}\ \emph {et~al.}(2017)\citenamefont
  {Van~Nieuwenburg}, \citenamefont {Liu},\ and\ \citenamefont
  {Huber}}]{van2017learning}%
  \BibitemOpen
  \bibfield  {author} {\bibinfo {author} {\bibfnamefont {E.~P.}\ \bibnamefont
  {Van~Nieuwenburg}}, \bibinfo {author} {\bibfnamefont {Y.-H.}\ \bibnamefont
  {Liu}},\ and\ \bibinfo {author} {\bibfnamefont {S.~D.}\ \bibnamefont
  {Huber}},\ }\href@noop {} {\bibfield  {journal} {\bibinfo  {journal} {Nature
  Physics}\ }\textbf {\bibinfo {volume} {13}},\ \bibinfo {pages} {435}
  (\bibinfo {year} {2017})}\BibitemShut {NoStop}%
\bibitem [{\citenamefont {Zhang}\ \emph {et~al.}(2019)\citenamefont {Zhang},
  \citenamefont {Liu},\ and\ \citenamefont {Wei}}]{zhang2019machine}%
  \BibitemOpen
  \bibfield  {author} {\bibinfo {author} {\bibfnamefont {W.}~\bibnamefont
  {Zhang}}, \bibinfo {author} {\bibfnamefont {J.}~\bibnamefont {Liu}},\ and\
  \bibinfo {author} {\bibfnamefont {T.-C.}\ \bibnamefont {Wei}},\ }\href@noop
  {} {\bibfield  {journal} {\bibinfo  {journal} {Physical Review E}\ }\textbf
  {\bibinfo {volume} {99}},\ \bibinfo {pages} {032142} (\bibinfo {year}
  {2019})}\BibitemShut {NoStop}%
\bibitem [{\citenamefont {Wang}(2016)}]{wang2016discovering}%
  \BibitemOpen
  \bibfield  {author} {\bibinfo {author} {\bibfnamefont {L.}~\bibnamefont
  {Wang}},\ }\href@noop {} {\bibfield  {journal} {\bibinfo  {journal} {Physical
  Review B}\ }\textbf {\bibinfo {volume} {94}},\ \bibinfo {pages} {195105}
  (\bibinfo {year} {2016})}\BibitemShut {NoStop}%
\bibitem [{\citenamefont {Shen}\ \emph
  {et~al.}(2021{\natexlab{a}})\citenamefont {Shen}, \citenamefont {Li},
  \citenamefont {Deng},\ and\ \citenamefont {Zhang}}]{shen2021supervised}%
  \BibitemOpen
  \bibfield  {author} {\bibinfo {author} {\bibfnamefont {J.}~\bibnamefont
  {Shen}}, \bibinfo {author} {\bibfnamefont {W.}~\bibnamefont {Li}}, \bibinfo
  {author} {\bibfnamefont {S.}~\bibnamefont {Deng}},\ and\ \bibinfo {author}
  {\bibfnamefont {T.}~\bibnamefont {Zhang}},\ }\href@noop {} {\bibfield
  {journal} {\bibinfo  {journal} {Physical Review E}\ }\textbf {\bibinfo
  {volume} {103}},\ \bibinfo {pages} {052140} (\bibinfo {year}
  {2021}{\natexlab{a}})}\BibitemShut {NoStop}%
\bibitem [{\citenamefont {Wetzel}(2017)}]{wetzel2017unsupervised}%
  \BibitemOpen
  \bibfield  {author} {\bibinfo {author} {\bibfnamefont {S.~J.}\ \bibnamefont
  {Wetzel}},\ }\href@noop {} {\bibfield  {journal} {\bibinfo  {journal}
  {Physical Review E}\ }\textbf {\bibinfo {volume} {96}},\ \bibinfo {pages}
  {022140} (\bibinfo {year} {2017})}\BibitemShut {NoStop}%
\bibitem [{\citenamefont {Hu}\ \emph {et~al.}(2017)\citenamefont {Hu},
  \citenamefont {Singh},\ and\ \citenamefont {Scalettar}}]{hu2017discovering}%
  \BibitemOpen
  \bibfield  {author} {\bibinfo {author} {\bibfnamefont {W.}~\bibnamefont
  {Hu}}, \bibinfo {author} {\bibfnamefont {R.~R.}\ \bibnamefont {Singh}},\ and\
  \bibinfo {author} {\bibfnamefont {R.~T.}\ \bibnamefont {Scalettar}},\
  }\href@noop {} {\bibfield  {journal} {\bibinfo  {journal} {Physical Review
  E}\ }\textbf {\bibinfo {volume} {95}},\ \bibinfo {pages} {062122} (\bibinfo
  {year} {2017})}\BibitemShut {NoStop}%
\bibitem [{\citenamefont {Wang}\ and\ \citenamefont
  {Zhai}(2017)}]{wang2017machine}%
  \BibitemOpen
  \bibfield  {author} {\bibinfo {author} {\bibfnamefont {C.}~\bibnamefont
  {Wang}}\ and\ \bibinfo {author} {\bibfnamefont {H.}~\bibnamefont {Zhai}},\
  }\href@noop {} {\bibfield  {journal} {\bibinfo  {journal} {Physical Review
  B}\ }\textbf {\bibinfo {volume} {96}},\ \bibinfo {pages} {144432} (\bibinfo
  {year} {2017})}\BibitemShut {NoStop}%
\bibitem [{\citenamefont {Wang}\ \emph {et~al.}(2021)\citenamefont {Wang},
  \citenamefont {Zhang}, \citenamefont {Hua},\ and\ \citenamefont
  {Wei}}]{wang2021unsupervised}%
  \BibitemOpen
  \bibfield  {author} {\bibinfo {author} {\bibfnamefont {J.}~\bibnamefont
  {Wang}}, \bibinfo {author} {\bibfnamefont {W.}~\bibnamefont {Zhang}},
  \bibinfo {author} {\bibfnamefont {T.}~\bibnamefont {Hua}},\ and\ \bibinfo
  {author} {\bibfnamefont {T.-C.}\ \bibnamefont {Wei}},\ }\href@noop {}
  {\bibfield  {journal} {\bibinfo  {journal} {Physical Review Research}\
  }\textbf {\bibinfo {volume} {3}},\ \bibinfo {pages} {013074} (\bibinfo {year}
  {2021})}\BibitemShut {NoStop}%
\bibitem [{\citenamefont {Pearson}(1901)}]{pearson1901liii}%
  \BibitemOpen
  \bibfield  {author} {\bibinfo {author} {\bibfnamefont {K.}~\bibnamefont
  {Pearson}},\ }\href@noop {} {\bibfield  {journal} {\bibinfo  {journal} {The
  London, Edinburgh, and Dublin Philosophical Magazine and Journal of Science}\
  }\textbf {\bibinfo {volume} {2}},\ \bibinfo {pages} {559} (\bibinfo {year}
  {1901})}\BibitemShut {NoStop}%
\bibitem [{\citenamefont {Abdi}\ and\ \citenamefont
  {Williams}(2010)}]{abdi2010principal}%
  \BibitemOpen
  \bibfield  {author} {\bibinfo {author} {\bibfnamefont {H.}~\bibnamefont
  {Abdi}}\ and\ \bibinfo {author} {\bibfnamefont {L.~J.}\ \bibnamefont
  {Williams}},\ }\href@noop {} {\bibfield  {journal} {\bibinfo  {journal}
  {Wiley interdisciplinary reviews: computational statistics}\ }\textbf
  {\bibinfo {volume} {2}},\ \bibinfo {pages} {433} (\bibinfo {year}
  {2010})}\BibitemShut {NoStop}%
\bibitem [{\citenamefont {Van~der Maaten}\ and\ \citenamefont
  {Hinton}(2008)}]{van2008visualizing}%
  \BibitemOpen
  \bibfield  {author} {\bibinfo {author} {\bibfnamefont {L.}~\bibnamefont
  {Van~der Maaten}}\ and\ \bibinfo {author} {\bibfnamefont {G.}~\bibnamefont
  {Hinton}},\ }\href@noop {} {\bibfield  {journal} {\bibinfo  {journal}
  {Journal of machine learning research}\ }\textbf {\bibinfo {volume} {9}}
  (\bibinfo {year} {2008})}\BibitemShut {NoStop}%
\bibitem [{\citenamefont {Van Der~Maaten}(2014)}]{van2014accelerating}%
  \BibitemOpen
  \bibfield  {author} {\bibinfo {author} {\bibfnamefont {L.}~\bibnamefont {Van
  Der~Maaten}},\ }\href@noop {} {\bibfield  {journal} {\bibinfo  {journal} {The
  Journal of Machine Learning Research}\ }\textbf {\bibinfo {volume} {15}},\
  \bibinfo {pages} {3221} (\bibinfo {year} {2014})}\BibitemShut {NoStop}%
\bibitem [{\citenamefont {Bourlard}\ and\ \citenamefont
  {Kamp}(1988)}]{bourlard1988auto}%
  \BibitemOpen
  \bibfield  {author} {\bibinfo {author} {\bibfnamefont {H.}~\bibnamefont
  {Bourlard}}\ and\ \bibinfo {author} {\bibfnamefont {Y.}~\bibnamefont
  {Kamp}},\ }\href@noop {} {\bibfield  {journal} {\bibinfo  {journal}
  {Biological cybernetics}\ }\textbf {\bibinfo {volume} {59}},\ \bibinfo
  {pages} {291} (\bibinfo {year} {1988})}\BibitemShut {NoStop}%
\bibitem [{\citenamefont {Hinton}\ and\ \citenamefont
  {Zemel}(1994)}]{hinton1994autoencoders}%
  \BibitemOpen
  \bibfield  {author} {\bibinfo {author} {\bibfnamefont {G.~E.}\ \bibnamefont
  {Hinton}}\ and\ \bibinfo {author} {\bibfnamefont {R.~S.}\ \bibnamefont
  {Zemel}},\ }\href@noop {} {\bibfield  {journal} {\bibinfo  {journal}
  {Advances in neural information processing systems}\ }\textbf {\bibinfo
  {volume} {6}},\ \bibinfo {pages} {3} (\bibinfo {year} {1994})}\BibitemShut
  {NoStop}%
\bibitem [{\citenamefont {Hinton}\ and\ \citenamefont
  {Salakhutdinov}(2006)}]{hinton2006reducing}%
  \BibitemOpen
  \bibfield  {author} {\bibinfo {author} {\bibfnamefont {G.~E.}\ \bibnamefont
  {Hinton}}\ and\ \bibinfo {author} {\bibfnamefont {R.~R.}\ \bibnamefont
  {Salakhutdinov}},\ }\href@noop {} {\bibfield  {journal} {\bibinfo  {journal}
  {science}\ }\textbf {\bibinfo {volume} {313}},\ \bibinfo {pages} {504}
  (\bibinfo {year} {2006})}\BibitemShut {NoStop}%
\bibitem [{\citenamefont {Shen}\ \emph
  {et~al.}(2021{\natexlab{b}})\citenamefont {Shen}, \citenamefont {Li},
  \citenamefont {Deng}, \citenamefont {Xu}, \citenamefont {Chen},\ and\
  \citenamefont {Liu}}]{shen2021machine}%
  \BibitemOpen
  \bibfield  {author} {\bibinfo {author} {\bibfnamefont {J.}~\bibnamefont
  {Shen}}, \bibinfo {author} {\bibfnamefont {W.}~\bibnamefont {Li}}, \bibinfo
  {author} {\bibfnamefont {S.}~\bibnamefont {Deng}}, \bibinfo {author}
  {\bibfnamefont {D.}~\bibnamefont {Xu}}, \bibinfo {author} {\bibfnamefont
  {S.}~\bibnamefont {Chen}},\ and\ \bibinfo {author} {\bibfnamefont
  {F.}~\bibnamefont {Liu}},\ }\href@noop {} {\bibinfo {title} {Machine learning
  of pair-contact process with diffusion}} (\bibinfo {year}
  {2021}{\natexlab{b}}),\ \Eprint {https://arxiv.org/abs/2112.00489}
  {arXiv:2112.00489 [cond-mat.stat-mech]} \BibitemShut {NoStop}%
\bibitem [{\citenamefont {Kouw}\ and\ \citenamefont
  {Loog}(2018)}]{kouw2018introduction}%
  \BibitemOpen
  \bibfield  {author} {\bibinfo {author} {\bibfnamefont {W.~M.}\ \bibnamefont
  {Kouw}}\ and\ \bibinfo {author} {\bibfnamefont {M.}~\bibnamefont {Loog}},\
  }\href@noop {} {\bibfield  {journal} {\bibinfo  {journal} {arXiv preprint
  arXiv:1812.11806}\ } (\bibinfo {year} {2018})}\BibitemShut {NoStop}%
\bibitem [{\citenamefont {Xu}\ \emph {et~al.}(2020)\citenamefont {Xu},
  \citenamefont {He},\ and\ \citenamefont {Shu}}]{xu2020transfer}%
  \BibitemOpen
  \bibfield  {author} {\bibinfo {author} {\bibfnamefont {W.}~\bibnamefont
  {Xu}}, \bibinfo {author} {\bibfnamefont {J.}~\bibnamefont {He}},\ and\
  \bibinfo {author} {\bibfnamefont {Y.}~\bibnamefont {Shu}},\ }\href@noop {}
  {\bibfield  {journal} {\bibinfo  {journal} {Advances and Applications in Deep
  Learning}\ ,\ \bibinfo {pages} {45}} (\bibinfo {year} {2020})}\BibitemShut
  {NoStop}%
\bibitem [{\citenamefont {Ch’Ng}\ \emph {et~al.}(2017)\citenamefont
  {Ch’Ng}, \citenamefont {Carrasquilla}, \citenamefont {Melko},\ and\
  \citenamefont {Khatami}}]{ch2017machine}%
  \BibitemOpen
  \bibfield  {author} {\bibinfo {author} {\bibfnamefont {K.}~\bibnamefont
  {Ch’Ng}}, \bibinfo {author} {\bibfnamefont {J.}~\bibnamefont
  {Carrasquilla}}, \bibinfo {author} {\bibfnamefont {R.~G.}\ \bibnamefont
  {Melko}},\ and\ \bibinfo {author} {\bibfnamefont {E.}~\bibnamefont
  {Khatami}},\ }\href@noop {} {\bibfield  {journal} {\bibinfo  {journal}
  {Physical Review X}\ }\textbf {\bibinfo {volume} {7}},\ \bibinfo {pages}
  {031038} (\bibinfo {year} {2017})}\BibitemShut {NoStop}%
\bibitem [{\citenamefont {Huembeli}\ \emph {et~al.}(2018)\citenamefont
  {Huembeli}, \citenamefont {Dauphin},\ and\ \citenamefont
  {Wittek}}]{huembeli2018identifying}%
  \BibitemOpen
  \bibfield  {author} {\bibinfo {author} {\bibfnamefont {P.}~\bibnamefont
  {Huembeli}}, \bibinfo {author} {\bibfnamefont {A.}~\bibnamefont {Dauphin}},\
  and\ \bibinfo {author} {\bibfnamefont {P.}~\bibnamefont {Wittek}},\
  }\href@noop {} {\bibfield  {journal} {\bibinfo  {journal} {Physical Review
  B}\ }\textbf {\bibinfo {volume} {97}},\ \bibinfo {pages} {134109} (\bibinfo
  {year} {2018})}\BibitemShut {NoStop}%
\bibitem [{\citenamefont {Malo~Roset}(2019)}]{malo2019applications}%
  \BibitemOpen
  \bibfield  {author} {\bibinfo {author} {\bibfnamefont {L.}~\bibnamefont
  {Malo~Roset}},\ }\emph {\bibinfo {title} {Applications of machine learning to
  studies of quantum phase transitions}},\ \href@noop {} {Master's thesis},\
  \bibinfo  {school} {Universitat Polit{\`e}cnica de Catalunya} (\bibinfo
  {year} {2019})\BibitemShut {NoStop}%
\bibitem [{\citenamefont {Eaton}\ and\ \citenamefont
  {Mansbach}(2012)}]{eaton2012spin}%
  \BibitemOpen
  \bibfield  {author} {\bibinfo {author} {\bibfnamefont {E.}~\bibnamefont
  {Eaton}}\ and\ \bibinfo {author} {\bibfnamefont {R.}~\bibnamefont
  {Mansbach}},\ }in\ \href@noop {} {\emph {\bibinfo {booktitle} {Proceedings of
  the AAAI Conference on Artificial Intelligence}}},\ Vol.~\bibinfo {volume}
  {26}\ (\bibinfo {year} {2012})\BibitemShut {NoStop}%
\bibitem [{\citenamefont {Zhang}\ \emph {et~al.}(2014)\citenamefont {Zhang},
  \citenamefont {Moore},\ and\ \citenamefont {Zdeborov{\'a}}}]{zhang2014phase}%
  \BibitemOpen
  \bibfield  {author} {\bibinfo {author} {\bibfnamefont {P.}~\bibnamefont
  {Zhang}}, \bibinfo {author} {\bibfnamefont {C.}~\bibnamefont {Moore}},\ and\
  \bibinfo {author} {\bibfnamefont {L.}~\bibnamefont {Zdeborov{\'a}}},\
  }\href@noop {} {\bibfield  {journal} {\bibinfo  {journal} {Physical Review
  E}\ }\textbf {\bibinfo {volume} {90}},\ \bibinfo {pages} {052802} (\bibinfo
  {year} {2014})}\BibitemShut {NoStop}%
\bibitem [{\citenamefont {Ajakan}\ \emph {et~al.}(2014)\citenamefont {Ajakan},
  \citenamefont {Germain}, \citenamefont {Larochelle}, \citenamefont
  {Laviolette},\ and\ \citenamefont {Marchand}}]{ajakan2014domain}%
  \BibitemOpen
  \bibfield  {author} {\bibinfo {author} {\bibfnamefont {H.}~\bibnamefont
  {Ajakan}}, \bibinfo {author} {\bibfnamefont {P.}~\bibnamefont {Germain}},
  \bibinfo {author} {\bibfnamefont {H.}~\bibnamefont {Larochelle}}, \bibinfo
  {author} {\bibfnamefont {F.}~\bibnamefont {Laviolette}},\ and\ \bibinfo
  {author} {\bibfnamefont {M.}~\bibnamefont {Marchand}},\ }\href@noop {}
  {\bibfield  {journal} {\bibinfo  {journal} {arXiv preprint arXiv:1412.4446}\
  } (\bibinfo {year} {2014})}\BibitemShut {NoStop}%
\bibitem [{\citenamefont {Hinrichsen}(2000)}]{hinrichsen2000non}%
  \BibitemOpen
  \bibfield  {author} {\bibinfo {author} {\bibfnamefont {H.}~\bibnamefont
  {Hinrichsen}},\ }\href@noop {} {\bibfield  {journal} {\bibinfo  {journal}
  {Advances in physics}\ }\textbf {\bibinfo {volume} {49}},\ \bibinfo {pages}
  {815} (\bibinfo {year} {2000})}\BibitemShut {NoStop}%
\bibitem [{\citenamefont {L{\"u}beck}(2004)}]{lubeck2004universal}%
  \BibitemOpen
  \bibfield  {author} {\bibinfo {author} {\bibfnamefont {S.}~\bibnamefont
  {L{\"u}beck}},\ }\href@noop {} {\bibfield  {journal} {\bibinfo  {journal}
  {International Journal of Modern Physics B}\ }\textbf {\bibinfo {volume}
  {18}},\ \bibinfo {pages} {3977} (\bibinfo {year} {2004})}\BibitemShut
  {NoStop}%
\bibitem [{\citenamefont {Ódor}(2004)}]{odor2004}%
  \BibitemOpen
  \bibfield  {author} {\bibinfo {author} {\bibfnamefont {G.}~\bibnamefont
  {Ódor}},\ }\href {https://doi.org/10.1103/revmodphys.76.663} {\bibfield
  {journal} {\bibinfo  {journal} {Reviews of Modern Physics}\ }\textbf
  {\bibinfo {volume} {76}},\ \bibinfo {pages} {663–724} (\bibinfo {year}
  {2004})}\BibitemShut {NoStop}%
\bibitem [{\citenamefont {Henkel}\ \emph {et~al.}(2008)\citenamefont {Henkel},
  \citenamefont {Hinrichsen}, \citenamefont {L{\"u}beck},\ and\ \citenamefont
  {Pleimling}}]{henkel2008non}%
  \BibitemOpen
  \bibfield  {author} {\bibinfo {author} {\bibfnamefont {M.}~\bibnamefont
  {Henkel}}, \bibinfo {author} {\bibfnamefont {H.}~\bibnamefont {Hinrichsen}},
  \bibinfo {author} {\bibfnamefont {S.}~\bibnamefont {L{\"u}beck}},\ and\
  \bibinfo {author} {\bibfnamefont {M.}~\bibnamefont {Pleimling}},\ }\href@noop
  {} {\emph {\bibinfo {title} {Non-equilibrium phase transitions}}},\
  Vol.~\bibinfo {volume} {1}\ (\bibinfo  {publisher} {Springer},\ \bibinfo
  {year} {2008})\BibitemShut {NoStop}%
\bibitem [{\citenamefont {Haken}(1975)}]{haken1975cooperative}%
  \BibitemOpen
  \bibfield  {author} {\bibinfo {author} {\bibfnamefont {H.}~\bibnamefont
  {Haken}},\ }\href@noop {} {\bibfield  {journal} {\bibinfo  {journal} {Reviews
  of Modern Physics}\ }\textbf {\bibinfo {volume} {47}},\ \bibinfo {pages} {67}
  (\bibinfo {year} {1975})}\BibitemShut {NoStop}%
\bibitem [{\citenamefont {Essam}(1980)}]{essam1980percolation}%
  \BibitemOpen
  \bibfield  {author} {\bibinfo {author} {\bibfnamefont {J.~W.}\ \bibnamefont
  {Essam}},\ }\href@noop {} {\bibfield  {journal} {\bibinfo  {journal} {Reports
  on progress in physics}\ }\textbf {\bibinfo {volume} {43}},\ \bibinfo {pages}
  {833} (\bibinfo {year} {1980})}\BibitemShut {NoStop}%
\bibitem [{\citenamefont {Christensen}\ and\ \citenamefont
  {Moloney}(2005)}]{christensen2005complexity}%
  \BibitemOpen
  \bibfield  {author} {\bibinfo {author} {\bibfnamefont {K.}~\bibnamefont
  {Christensen}}\ and\ \bibinfo {author} {\bibfnamefont {N.~R.}\ \bibnamefont
  {Moloney}},\ }\href@noop {} {\emph {\bibinfo {title} {Complexity and
  criticality}}},\ Vol.~\bibinfo {volume} {1}\ (\bibinfo  {publisher} {World
  Scientific Publishing Company},\ \bibinfo {year} {2005})\BibitemShut
  {NoStop}%
\bibitem [{\citenamefont {Jensen}\ and\ \citenamefont
  {Dickman}(1993)}]{jensen1993nonequilibrium}%
  \BibitemOpen
  \bibfield  {author} {\bibinfo {author} {\bibfnamefont {I.}~\bibnamefont
  {Jensen}}\ and\ \bibinfo {author} {\bibfnamefont {R.}~\bibnamefont
  {Dickman}},\ }\href@noop {} {\bibfield  {journal} {\bibinfo  {journal}
  {Physical Review E}\ }\textbf {\bibinfo {volume} {48}},\ \bibinfo {pages}
  {1710} (\bibinfo {year} {1993})}\BibitemShut {NoStop}%
\bibitem [{\citenamefont {Rossi}\ \emph {et~al.}(2000)\citenamefont {Rossi},
  \citenamefont {Pastor-Satorras},\ and\ \citenamefont
  {Vespignani}}]{rossi2000universality}%
  \BibitemOpen
  \bibfield  {author} {\bibinfo {author} {\bibfnamefont {M.}~\bibnamefont
  {Rossi}}, \bibinfo {author} {\bibfnamefont {R.}~\bibnamefont
  {Pastor-Satorras}},\ and\ \bibinfo {author} {\bibfnamefont {A.}~\bibnamefont
  {Vespignani}},\ }\href@noop {} {\bibfield  {journal} {\bibinfo  {journal}
  {Physical review letters}\ }\textbf {\bibinfo {volume} {85}},\ \bibinfo
  {pages} {1803} (\bibinfo {year} {2000})}\BibitemShut {NoStop}%
\bibitem [{\citenamefont {Ganin}\ \emph {et~al.}(2016)\citenamefont {Ganin},
  \citenamefont {Ustinova}, \citenamefont {Ajakan}, \citenamefont {Germain},
  \citenamefont {Larochelle}, \citenamefont {Laviolette}, \citenamefont
  {Marchand},\ and\ \citenamefont {Lempitsky}}]{ganin2016domain}%
  \BibitemOpen
  \bibfield  {author} {\bibinfo {author} {\bibfnamefont {Y.}~\bibnamefont
  {Ganin}}, \bibinfo {author} {\bibfnamefont {E.}~\bibnamefont {Ustinova}},
  \bibinfo {author} {\bibfnamefont {H.}~\bibnamefont {Ajakan}}, \bibinfo
  {author} {\bibfnamefont {P.}~\bibnamefont {Germain}}, \bibinfo {author}
  {\bibfnamefont {H.}~\bibnamefont {Larochelle}}, \bibinfo {author}
  {\bibfnamefont {F.}~\bibnamefont {Laviolette}}, \bibinfo {author}
  {\bibfnamefont {M.}~\bibnamefont {Marchand}},\ and\ \bibinfo {author}
  {\bibfnamefont {V.}~\bibnamefont {Lempitsky}},\ }\href@noop {} {\bibfield
  {journal} {\bibinfo  {journal} {The journal of machine learning research}\
  }\textbf {\bibinfo {volume} {17}},\ \bibinfo {pages} {2096} (\bibinfo {year}
  {2016})}\BibitemShut {NoStop}%
\bibitem [{\citenamefont {Kingma}\ and\ \citenamefont
  {Ba}(2017)}]{kingma2017adam}%
  \BibitemOpen
  \bibfield  {author} {\bibinfo {author} {\bibfnamefont {D.~P.}\ \bibnamefont
  {Kingma}}\ and\ \bibinfo {author} {\bibfnamefont {J.}~\bibnamefont {Ba}},\
  }\href@noop {} {\bibinfo {title} {Adam: A method for stochastic
  optimization}} (\bibinfo {year} {2017}),\ \Eprint
  {https://arxiv.org/abs/1412.6980} {arXiv:1412.6980 [cs.LG]} \BibitemShut
  {NoStop}%
\bibitem [{\citenamefont {Wetzel}\ and\ \citenamefont
  {Scherzer}(2017)}]{wetzel2017machine}%
  \BibitemOpen
  \bibfield  {author} {\bibinfo {author} {\bibfnamefont {S.~J.}\ \bibnamefont
  {Wetzel}}\ and\ \bibinfo {author} {\bibfnamefont {M.}~\bibnamefont
  {Scherzer}},\ }\href@noop {} {\bibfield  {journal} {\bibinfo  {journal}
  {Physical Review B}\ }\textbf {\bibinfo {volume} {96}},\ \bibinfo {pages}
  {184410} (\bibinfo {year} {2017})}\BibitemShut {NoStop}%
\bibitem [{\citenamefont {D'Angelo}\ and\ \citenamefont
  {B\"ottcher}(2020)}]{PhysRevResearch.2.023266}%
  \BibitemOpen
  \bibfield  {author} {\bibinfo {author} {\bibfnamefont {F.}~\bibnamefont
  {D'Angelo}}\ and\ \bibinfo {author} {\bibfnamefont {L.}~\bibnamefont
  {B\"ottcher}},\ }\href {https://doi.org/10.1103/PhysRevResearch.2.023266}
  {\bibfield  {journal} {\bibinfo  {journal} {Phys. Rev. Research}\ }\textbf
  {\bibinfo {volume} {2}},\ \bibinfo {pages} {023266} (\bibinfo {year}
  {2020})}\BibitemShut {NoStop}%
\end{thebibliography}%

\end{document}